\documentclass{aa}
\usepackage{dsfont}   
\usepackage{float}
\usepackage{booktabs}
\begin{document}

\title{MAHA-z: A Mahalanobis distance $k$NN estimator for photometric redshifts}

\author{Paula S. Ferreira\inst{1,2} \and
        Ribamar R. R. Reis\inst{3,4} \and
        Renyue Cen\inst{1,2}
}

\institute{Center for Cosmology and Computational Astrophysics, Institute for Advanced Study in Physics, Zhejiang University, Hangzhou 310058, China\\
\email{ferreira\_paula@zju.edu.cn}
\and
Institute of Astronomy, School of Physics, Zhejiang University, Hangzhou 310058, China\\
\email{renyuecen@zju.edu.cn}
\and
Instituto de F\'isica, Universidade Federal do Rio de Janeiro, Av. Athos da Silveira Ramos, 149 -- Cidade Universit\'aria, Rio de Janeiro, Brasil
\and
Observat\'orio do Valongo, Universidade Federal do Rio de Janeiro, Ladeira Pedro Ant\^onio, 43, Centro, Rio de Janeiro, Brasil\\
\email{ribamar@if.ufrj.br}
}

\date{Received ; accepted }

\abstract
{We present \texttt{MAHA-z}, a photometric-redshift (photo-$z$) estimator that combines $k$-nearest neighbour ($k$NN) regression with a Mahalanobis distance metric learned from the covariance of a spectroscopic reference sample. For each target galaxy, \texttt{MAHA-z} performs a neighbour search in a colour-magnitude feature space with optional morphological parameters, and returns both a point estimate and a redshift posterior $p(z\mid\mathbf{f})$ obtained by kernel density estimation over the neighbour redshifts. The Mahalanobis metric accounts for correlated features and heterogeneous units, eliminating ad hoc normalisation, and its subspace restriction allows \texttt{MAHA-z} to handle missing bands on a per-galaxy basis without imputation. We validate the method on four datasets spanning simulations and survey photometry: the Euclid Flagship~2 mock, the Legacy-KiDZ calibration sample, VIPERS, and the DES Year~3 BAO sample. On Euclid Flagship~2 (LSST+NISP) we achieve $\sigma_{\rm NMAD}=0.0148$ with a $0.27\%$ outlier fraction, and on DES Y3 we obtain $\sigma_{\rm NMAD}\approx0.023$ with $<1\%$ outliers and posterior calibration comparable to the \texttt{ANNz2} catalogue. \texttt{MAHA-z} processes millions of galaxies in minutes and provides competitive PDFs at substantially lower computational cost than trained machine-learning models, making it a practical option for iterative photo-$z$ production and validation in Stage-IV surveys.}

\titlerunning{MAHA-z}
\authorrunning{P.S. Ferreira et al.}

\maketitle

\section{Introduction}
\label{sec:intro}

Photometric redshift (photo-z) estimation is a key technique for large-scale structure (LSS) surveys: multi-band imaging is quicker and yields far more detected objects than spectroscopic observations. Upcoming projects such as \textit{Euclid} \citep{euclid}, the Vera C. Rubin Observatory (LSST) \citep{abell2009lsst}, the China Space Station Telescope (CSST) \citep{csst}, and the Nancy Grace Roman Space Telescope \citep{bartusek2022nancy} are expected to observe billions of galaxies. It is therefore important to benchmark and validate the many photo-z estimation codes that may be applied to these surveys, especially when the goal is to find faint objects.

Template-fitting codes compare multi-band photometry with libraries of galaxy spectral templates, often within a Bayesian or chi-squared framework, and can provide full redshift probability density functions (PDFs); they are physically motivated and can handle faint or unusual objects, but depend on template completeness and calibration. Machine-learning codes instead learn a mapping from photometry to redshift using a spectroscopic training sample; once trained they are extremely fast and scalable to huge catalogues, but their accuracy depends on the representativeness of the training data and on robust uncertainty estimation. Both families are essential for Stage-IV surveys, and the codes below represent well-tested implementations that have been widely used in survey pipelines and photo-z validation.

\texttt{LePhare} \citep{arnouts1999measuring,guzzo2014vimos} is a versatile template-fitting code that performs spectral energy distribution (SED) fitting against libraries of stellar, galaxy, and active galactic nucleus (AGN) templates. Its importance lies in its flexibility: it provides photo-z estimates together with physical parameters such as stellar mass and star-formation history. Because it can handle non-stellar SEDs, \texttt{LePhare} is also valuable for AGN photo-$z$ studies and for cross-checks of galaxy photo-$z$ catalogues. \texttt{BPZ} \citep{benitez2000bayesian} is a Bayesian template-fitting code that combines observed photometry with a prior on the redshift distribution to estimate photo-$z$ PDFs. Its importance stems from its probabilistic output and its long-standing validation at faint magnitudes with filter sets similar to those of current and near-future surveys. \texttt{BPZ} has been adopted in major survey analyses and remains a standard template-fitting reference against which newer methods are compared. \texttt{ZEBRA} \citep{feldmann2006zurich} is also a Bayesian template-fitting code also designed for SED analysis. It provides a well-defined statistical framework for combining template fits with priors and for deriving redshift PDFs, making it useful for validating photo-$z$ systematics and for applications where a Bayesian template-fitting approach is preferred. \texttt{EAZY} \citep{brammer2008eazy} is a fast and flexible template-fitting photo-$z$ code that uses a configurable template set and efficient fitting procedures. Its importance is practical: it is computationally efficient for large catalogues, widely used in deep-field and survey analyses, and provides redshift PDFs that can be propagated into cosmological analyses. \texttt{EAZY} is particularly useful when rapid, flexible template-fitting estimates are needed across heterogeneous data sets.

\texttt{TPZ} \citep{10.1093/mnras/stt574} is a machine-learning photo-$z$ code based on tree-based methods, providing redshift estimates and PDFs from photometric data. It offers a non-linear, training-based alternative to template fitting and has been used to demonstrate that machine-learning methods can deliver competitive photo-$z$ performance with well-characterised uncertainties. \texttt{ANNz2} \citep{sadeh2016annz2} is a machine-learning photo-$z$ implementation that uses artificial neural networks and boosted decision trees and can produce full redshift PDFs. \texttt{ANNz2} can handle large feature sets, provide probabilistic outputs, and serve as a strong machine-learning benchmark for survey photo-$z$ catalogues. \texttt{GPZ} \citep{almosallam2016gpz} is a Gaussian-process-based machine-learning code for photo-$z$ estimation, it provides non-linear predictions together with principled uncertainty estimates, making it well suited to propagating photo-$z$ errors and to identifying objects where the training-based mapping is less reliable. \texttt{DNF} \citep{de2016dnf} is a based on directional neighbourhood fitting that has been extensively used in the Dark Energy Survey \citep{Toribio_San_Cipriano_2024}. It provides robust photo-$z$ PDFs, performs well on deep survey data, and has been validated as a machine-learning counterpart to template-fitting codes in large photometric catalogues, making it suitable for future LSST- and \textit{Euclid}-like analyses.

It should be noted that template-matching codes such as \texttt{LePhare}, \texttt{BPZ}, \texttt{ZEBRA}, and \texttt{EAZY} depend on the adopted template library and on prior choices, can suffer from colour-redshift degeneracies, and may be biased for objects whose SEDs are poorly represented by the templates, including strong emission-line galaxies and AGN \citep{Mobasher_2007}. Quasars add a further complication: their point-like appearance and colours often mimic those of stars, and such misclassifications propagate into photo-$z$ errors and bias cosmological inference; recent machine-learning work mitigates this by using a K-Nearest Neighbours classifier to separate quasars from stellar contaminants at $0.99$ precision and a hybrid \texttt{ANNz2}/decision-tree photo-$z$ pipeline with outlier suppression, producing a cleaned DES quasar catalogue suitable for large-scale structure studies out to $z\approx3$ and a full sample reliable to $z\approx4$ \citep{motta2026photometricclassificationquasarsdes}. They can also be computationally expensive for very large catalogues, although \texttt{EAZY} mitigates this to some extent, and Bayesian priors in \texttt{BPZ} and \texttt{ZEBRA} can introduce biases when mismatched to the data. Machine-learning codes such as \texttt{TPZ}, \texttt{ANNz2}, \texttt{GPZ}, and \texttt{DNF} require a representative spectroscopic training sample, can extrapolate poorly to faint or rare objects outside the training distribution, and may produce overconfident PDFs if the training data are not fully representative; their performance is sensitive to feature selection and training choices, Gaussian processes in \texttt{GPZ} scale poorly with very large training sets, and nearest-neighbour methods such as \texttt{DNF} depend on the density and coverage of the training sample. Consequently, no single code is universally optimal, and robust photo-z catalogues for Stage-IV surveys rely on combining and cross-validating template-fitting and machine-learning estimates.

\cite{Abdalla_2011} compared several widely used codes, including \texttt{ANNz2}, \texttt{HyperZ} \citep{bolzonella2000photometricredshiftsbasedstandard}, SDSS, \texttt{LePHARE}, \texttt{BPZ}, and \texttt{ZEBRA}, on a large sample of luminous red galaxies and found that all methods exhibit redshift-dependent biases. In their analysis, template-fitting codes that use updated observed templates, particularly \texttt{LePHARE}, perform best in the lower-redshift bins, whereas the training-based \texttt{ANNz2} performs best at intermediate redshifts where the spectroscopic training set is complete and contains many training objects. At higher redshifts, or in regimes where the training sample is sparse or unrepresentative, machine-learning methods degrade and template-fitting methods can suffer from colour--redshift degeneracies and template mismatch; the ranking between methods can therefore reverse. This redshift dependence is a central reason why photo-z catalogues for future surveys rely on combining and cross-validating template-fitting and machine-learning codes such as \texttt{BPZ}, \texttt{ANNz}, \texttt{DNF}, and \texttt{LePHARE}, rather than adopting a single algorithm for the whole survey \citep{Ferreira:2025mnb}.

Another approach, the nearest-neighbour regression that dominated photo-z estimation before the template and neural-network era, avoids both limitations but has long been considered too imprecise, because it treats every direction in colour space as equally meaningful. The Euclidean metric at the heart of standard $k$NN ignores the fact that the galaxy population occupies an elongated, correlated locus in colour--magnitude space, and it consequently weights differences along
the main colour sequence in the same way as differences perpendicular to it, even though the latter are far more informative about redshift.

In this paper we revisit the $k$NN approach with a metric that respects the geometry of the data. The idea of using the local neighbourhood of a target galaxy to estimate its photo-$z$ uncertainty, rather than relying on global training residuals, was introduced by the Nearest Neighbor Error (NNE) method of \citet{Oyaizu_2008} and subsequently refined by \cite{Reis_2012} for the SDSS Coadd photo-$z$ catalogue. The refinement addresses a subtle bias in the original formulation: because the ANN photo-$z$ values of training galaxies are over-fitted to their spectroscopic labels, the neighbour scatter computed within the training set underestimates the true error of an unseen photometric target. \cite{Reis_2012} instead use an independent validation set as the neighbour pool for NNE, so that the neighbour redshifts reflect the same out-of-sample accuracy as the target galaxies. \cite{Reis_2012} also pioneered in including morphological features to the photo-z estimation and found that the improvement was not significant.

\texttt{MAHA-z} builds directly on that idea and extends it in two directions. First, it uses the same neighbourhood for the full estimate rather than for the error alone: the point estimate is the neighbour mean, the per-galaxy uncertainty is the neighbour scatter, and the posterior is a kernel density estimate over the neighbour redshifts. This replaces the ANN point estimate of the NNE approach and is what makes a full $p(z \mid \mathbf{f})$ possible. Second, it replaces the Euclidean distance with the Mahalanobis distance \citep{mahalanobis}\footnote{\textbf{Prasanta Chandra Mahalanobis} (1893--1972) was an Indian statistician who spent most of his working life in Calcutta, where he founded the Indian Statistical Institute in 1931 and helped turn large-scale sample surveys into a practical tool.
The distance that carries his name came out of his 1936 paper \emph{On the generalised distance in statistics}, where he introduced it as the $D^{2}$ statistic.}, estimated from the covariance matrix of the spectroscopic training set, so that the neighbourhood respects the elongated, correlated locus of the galaxy population rather than treating every direction in colour space as equally meaningful. A further consequence of the Mahalanobis formulation is that missing photometry is handled natively, because the sub-block of the precision matrix corresponding to the available features is itself the correct metric in that subspace.

\section{The \texttt{MAHA-z} photo-$z$ estimator}
\label{sec:mahaz}

The Mahalanobis distance \citep{mahalanobis} generalises the Euclidean distance to a space whose coordinates are correlated and carry different physical units. Given a random vector $\mathbf{x} \in \mathbb{R}^{d}$ with mean $\boldsymbol{\mu}$ and covariance matrix $\mathbf{C} = \mathbb{E}\bigl[(\mathbf{x}-\boldsymbol{\mu})(\mathbf{x}-\boldsymbol{\mu})^{\top}\bigr]$, the squared Mahalanobis distance of a point $\mathbf{x}$ from the distribution is
\begin{equation}
    d^{2}(\mathbf{x}) =
    (\mathbf{x} - \boldsymbol{\mu})^{\!\top}\,
    \mathbf{C}^{-1}\,
    (\mathbf{x} - \boldsymbol{\mu}),
    \label{eq:mahalanobis_intro}
\end{equation}
and, between two arbitrary points $\mathbf{x}$ and $\mathbf{y}$,
\begin{equation}
    d^{2}(\mathbf{x}, \mathbf{y}) =
    (\mathbf{x} - \mathbf{y})^{\!\top}\,
    \mathbf{C}^{-1}\,
    (\mathbf{x} - \mathbf{y}).
    \label{eq:mahalanobis_pair}
\end{equation}

The inverse covariance matrix $\mathbf{C}^{-1}$ acts as a metric tensor: it rescales each coordinate by its own variance and rotates the axes along the principal directions of the distribution, so that the unit ball of the metric---the locus $d^{2}(\mathbf{x}) = 1$---is an ellipsoid whose axes are the eigenvectors of $\mathbf{C}$ and whose semiaxes are the corresponding standard deviations $\sqrt{\lambda_{i}}$. Equivalently, if $\mathbf{x} \sim \mathcal{N}(\boldsymbol{\mu}, \mathbf{C})$ then $d^{2}(\mathbf{x})$ follows a $\chi^{2}$ distribution with $d$ degrees of freedom, which gives the metric a probabilistic interpretation: a point at Mahalanobis distance $d$ lies on the $d$-sigma contour of the distribution, independent of the number of dimensions and of the correlation between them. Two properties are what motivate its use here. First, unlike the Euclidean metric, it is invariant under
invertible linear transformations of the coordinates, so a rescaling or rotation of the feature space leaves every distance unchanged -- an essential property when the coordinates are magnitudes and colours measured in different bands with very different variances. Second, it reduces to the Euclidean distance in the special case $\mathbf{C} = \mathbf{I}$, so it is a strict generalisation rather than a different metric, and the ``information content'' of the covariance matrix can be switched off for diagnostic purposes.

We propose to use this method to estimate photometric redshift. \texttt{MAHA-z}\footnote{The code will be available at \url{https://github.com/psilvaf/MAHA-z/} after publication.} assigns a photometric redshift to each target galaxy by combining three ingredients: (i) a feature vector built from magnitudes, colours, and optionally morphological parameters; (ii) a Mahalanobis metric whose covariance encodes the structure of the galaxy population; and (iii) a $k$-nearest neighbour regression whose mean gives the point estimate and whose spread gives the per-galaxy uncertainty. A kernel density estimate (KDE) over the same neighbourhood optionally provides a full posterior $p(z \mid \mathbf{f})$. Each ingredient, and the exact configuration used in the reference implementation, is described below.

\subsection{Feature space}
\label{sec:feature_space}

For each galaxy the feature vector is
\begin{equation}
    \mathbf{f} =
    \bigl(
        m_1, \ldots, m_{N_b},\;
        c_1, \ldots, c_{N_c},\;
        \mu_1, \ldots, \mu_{N_m}
    \bigr),
    \label{eq:feature_vector}
\end{equation}
where $m_i$ are observed magnitudes, $c_j = m_a - m_b$ are colours, and $\mu_k$ are optional morphological parameters (e.g.\ S\'ersic index, effective radius, bulge fraction) included when available. Including both magnitudes and colours is deliberate: colours are more sensitive to redshift, while magnitudes provide a luminosity anchor that helps break colour--redshift degeneracies.

Three colour configurations supported are:
\begin{itemize}
    \item \texttt{all} (default): all pairwise colours,
          $N_c = \binom{N_b}{2}$;
    \item \texttt{adjacent}: consecutive colours,
          $N_c = N_b - 1$;
    \item \texttt{reference}: each band against a single reference band,
          $N_c = N_b - 1$;
    \item \texttt{none}: no colours, only magnitudes, $N_c = 0$.
\end{itemize}

\subsection{Covariance matrix}
\label{sec:covariance}

The covariance matrix $\mathbf{C}$ used by the Mahalanobis metric is estimated from the spectroscopic \emph{training} sample, with the mean photometric error variance of the \emph{photometric} sample added on the diagonal:
\begin{equation}
\mathbf{C} =
\begin{bmatrix}
\sigma_{11}^2 + \langle \epsilon_{1}^2 \rangle & \sigma_{12}^2 & \cdots & \sigma_{1d}^2 \\
\sigma_{21}^2 & \sigma_{22}^2 + \langle \epsilon_{2}^2 \rangle & \cdots & \sigma_{2d}^2 \\
\vdots & \vdots & \ddots & \vdots \\
\sigma_{d1}^2 & \sigma_{d2}^2 & \cdots & \sigma_{dd}^2 + \langle \epsilon_{d}^2 \rangle
\end{bmatrix} + \delta \mathbf{I},
\label{eq:cov_full}
\end{equation}
where $\sigma_{ij}^2$ is the covariance of the training features, $\langle \epsilon_i^2 \rangle$ is the mean squared $1\sigma$ photometric error of feature $i$ over the photometric sample. If the training sample is small, the empirical covariance can optionally be shrunk towards a scaled identity via the Ledoit--Wolf estimator; the two estimates are averaged with equal weight.

The squared Mahalanobis distance between a target galaxy $a$ and a spectroscopic galaxy $s$ is
\begin{equation}
    d^2_{a,s}
    = (\mathbf{f}_a - \mathbf{f}_s)^{\!\top}\,
      \mathbf{C}^{-1}\,
      (\mathbf{f}_a - \mathbf{f}_s),
    \label{eq:mahalanobis}
\end{equation}
and is evaluated with the pseudo-inverse of $\mathbf{C}$.

\subsection{$k$-nearest neighbour regression}
\label{sec:knn}

For each target galaxy $\mathbf{f}_a$, we select the $k$ training galaxies with the smallest distance $d^2_{a,s}$,
\begin{equation}
    \mathcal{N}_k(a) = \bigl\{ s_1, \ldots, s_k \bigr\},
    \qquad
    s_i = \operatorname*{arg\,min}^{(i)}_s d^2_{a,s},
    \label{eq:mahaz_neigh}
\end{equation}
and estimate the photometric redshift as the mean of their spectroscopic redshifts,
\begin{equation}
    z_{\rm phot}(a)
    = \frac{1}{k} \sum_{s \in \mathcal{N}_k(a)} z_{{\rm spec},s}.
    \label{eq:zphot_mean}
\end{equation}
The corresponding per-galaxy $1\sigma$ uncertainty is the standard deviation of the same neighbourhood,
\begin{align}
    \sigma_{z,a}
    = \sqrt{\frac{1}{k-1}\sum_{s \in \mathcal{N}_k(a)}
        \bigl(z_{{\rm spec},s} - \bar{z}_a\bigr)^{2}},
    \\
    \bar{z}_a = \frac{1}{k}\sum_{s \in \mathcal{N}_k(a)} z_{{\rm spec},s}.
    \label{eq:sigma_za}
\end{align}

\begin{figure}
    \centering
    \includegraphics[width=\linewidth]{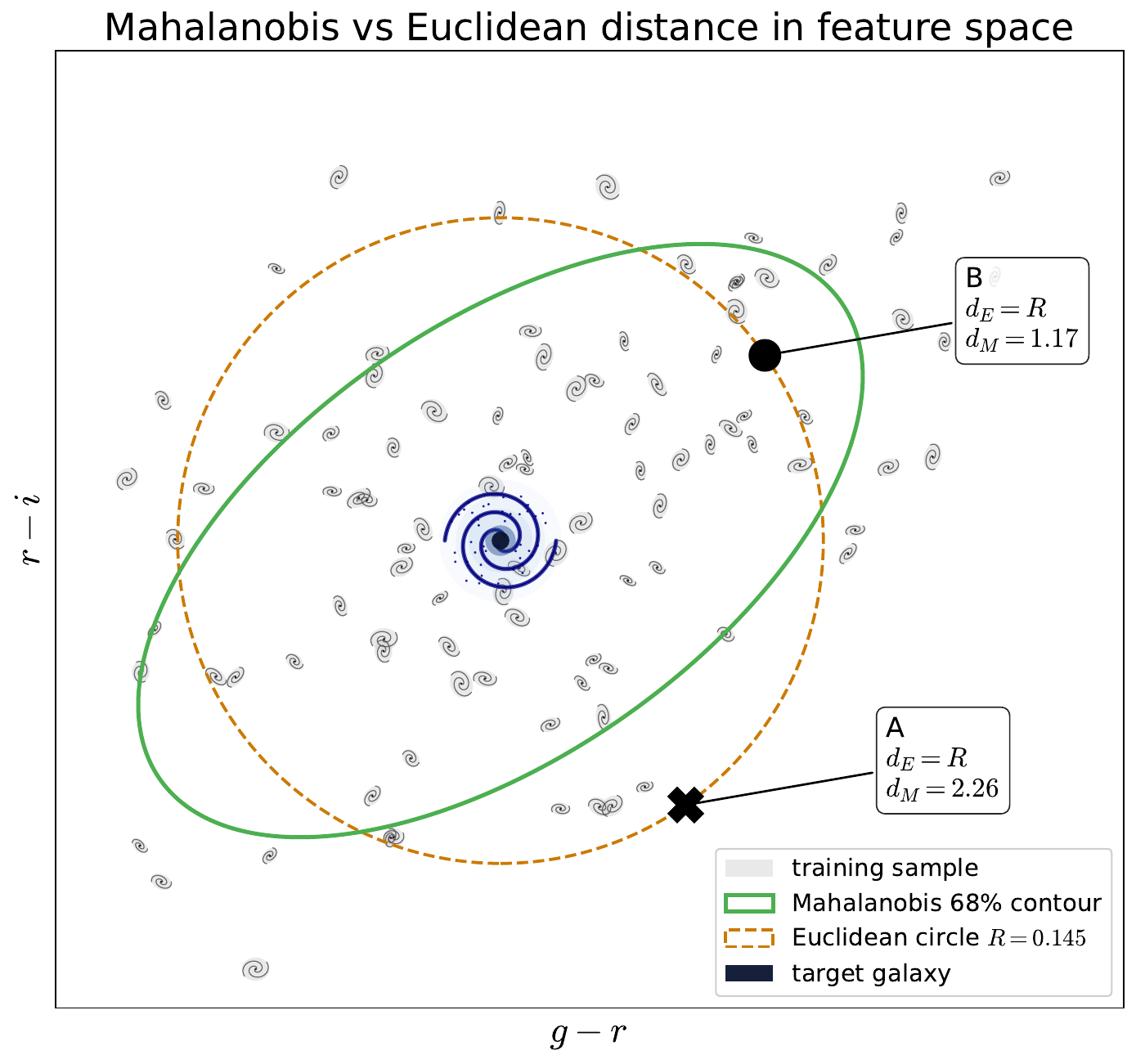}
    \caption{Mahalanobis versus Euclidean distance in the two-dimensional
    feature space of colours $g-r$ and $r-i$. Grey points: training
    galaxies, tracing an elongated, correlated locus. The green solid
    ellipse marks the 68\% Mahalanobis contour of the target galaxy;
    the orange dashed circle marks the set of points at the same
    Euclidean distance $R = 1.5\,\sigma_{\rm minor}$. Galaxy~A (minor
    axis, anomalous colour) and galaxy~B (major axis, typical colour)
    sit on this circle, yet their Mahalanobis distances differ by a
    factor of $\sim 2$ ($d_M = 2.27$ vs $d_M = 1.17$).}
    \label{fig:mahaz_schematic}
\end{figure}

Figure~\ref{fig:mahaz_schematic} illustrates why the Mahalanobis metric differs from a Euclidean one in a correlated colour space. The training galaxies (grey galaxies) trace an elongated locus. The green solid ellipse marks the 68\% Mahalanobis contour around a target galaxy; the orange dashed circle marks the set of points at the same Euclidean distance $R = 1.5\,\sigma_{\rm minor}$ from the target. Two training galaxies sit on this circle: galaxy~A lies along the minor axis
(an anomalous colour, off the locus) and galaxy~B along the major axis (a typical colour, on the locus). Although both are equidistant from the target in Euclidean terms, the Mahalanobis metric ranks~B a factor of $\sim 2$ closer than~A ($d_M = 1.17$ vs $d_M = 2.27$), because it measures distance in units of the local scatter of the galaxy population. The metric therefore weights neighbours according to the structure of the population rather than treating every direction in colour space as equally meaningful.

\subsection{Heterogeneous photometry: Mahalanobis in a feature subspace}
\label{sec:subspace}

A key property of the Mahalanobis metric is that it can be evaluated in any subset of features without imputation. For a target galaxy with a mask $m$ (the set of features that are finite for that galaxy), the sub-covariance is the corresponding sub-block of the full matrix,
\begin{equation}
    \mathbf{C}_m = \mathbf{C}[m, m],
    \qquad
    \mathbf{W}_m = \bigl(\mathbf{C}_m + \delta \mathbf{I}\bigr)^{-1},
    \label{eq:subspace}
\end{equation}
and the squared Mahalanobis distance in the subspace is
\begin{equation}
    d^2_m(\mathbf{x}, \mathbf{y})
    = (\mathbf{x}_m - \mathbf{y}_m)^{\!\top}\,
      \mathbf{W}_m\,
      (\mathbf{x}_m - \mathbf{y}_m).
    \label{eq:mahalanobis_subspace}
\end{equation}
This is mathematically consistent because the marginal distribution of a multivariate Gaussian over any subset of coordinates is Gaussian, with covariance equal to the corresponding sub-block of $\mathbf{C}$. The property is not shared by neural-network-based estimators, which require a fixed input dimensionality.

In the reference implementation this property is activated by the switch \texttt{use\_available\_bands}. Targets are grouped by the mask $m$ they carry, and each group is regressed against the training galaxies that are complete in that mask, provided at least \texttt{min\_training\_for\_mask} such galaxies exist; otherwise the group is left un-estimated (its $z_{\rm phot}$ is stored as NaN). This makes \texttt{MAHA-z} directly applicable to surveys with heterogeneous
photometric coverage.

\subsection{Photometric redshift probability density functions}
\label{sec:pdf_method}

The estimator can optionally return a full posterior $p(z \mid \mathbf{f})$, built as a Gaussian kernel density estimate over the redshifts of the $k$ neighbours selected in Eq.~\eqref{eq:mahaz_neigh}. We stress that the \emph{point estimate} is the neighbour mean of Eq.~\eqref{eq:zphot_mean}, not the mode of this KDE; the KDE is a complementary output for downstream applications that need a full posterior.

The kernel bandwidth is chosen adaptively per galaxy,
\begin{equation}
    h_a = \max\!\Bigl[\,\alpha \bigl(1 + \tilde{z}_a\bigr),\; h_{\min}\,\Bigr].
    \label{eq:bandwidth}
\end{equation}

The PDF is evaluated on a fixed grid
\begin{equation}
    z \in [0, z_{\rm max}], \qquad N_z = 100,
    \label{eq:zgrid}
\end{equation}
chosen wide enough to cover the typical survey range and uniform across all galaxies, so that downstream stacks and quantiles are directly comparable. The density is
\begin{align}
    p(z \mid \mathbf{f}_a)
    = \frac{1}{|\mathcal{N}_k(a)|}
      \sum_{s \in \mathcal{N}_k(a)}
      \frac{1}{\sqrt{2\pi}\,h_a}\\ \nonumber
      \exp\!\left[-\frac{1}{2}
        \left(\frac{z - z_{{\rm spec},s}}{h_a}\right)^{\!2}\right],
    \label{eq:kde}
\end{align}
and is renormalised to unit integral on the grid.

For every galaxy with a PDF, the following columns are written alongside the point estimate:
\begin{itemize}
    \item the mode, $z_{\rm mode} = \arg\max_z p(z \mid \mathbf{f}_a)$;
    \item the median, $z_{\rm med}$, defined by
          $\int_{-\infty}^{z_{\rm med}} p\,dz = 0.50$;
    \item the 68\% credible interval
          $[z_{16}, z_{84}]$, defined by
          $\int_{-\infty}^{z_{16}} p\,dz = 0.16$ and
          $\int_{-\infty}^{z_{84}} p\,dz = 0.84$;
    \item the width $\sigma_{68} = (z_{84} - z_{16})/2$.
\end{itemize}
These quantities are derived from the same neighbour set as the point estimate, so they are internally consistent with
Eqs.~\eqref{eq:zphot_mean}--\eqref{eq:sigma_za}.

\subsection{Training / validation split}
\label{sec:split}

Two modes are supported:

\begin{itemize}
    \item \textbf{Simulation-like mode} (\texttt{use\_all\_spec\_z=False}, default): a fraction \texttt{frac\_spec} of the catalogue is randomly drawn as the spectroscopic pool, and a fraction \texttt{val\_frac}     of that pool is held out for validation.
    \item \textbf{Real-data mode} (\texttt{use\_all\_spec\_z=True}): every row with a finite spectroscopic redshift enters the pool, and a fraction \texttt{val\_frac} of it is held out for validation;
    \texttt{frac\_spec} is ignored. Rows with NaN spectroscopic redshift are treated as photometric targets.
\end{itemize}

\section{Validation metrics}
\label{sec:metrics}

To quantify the performance of the estimator, we compare $z_{\rm phot}$ against $z_{\rm spec}$ on the held-out validation sample. Following the photo-$z$ literature we adopt the normalised residual
\begin{equation}
    \Delta z = \frac{z_{\rm phot} - z_{\rm spec}}{1 + z_{\rm spec}},
\end{equation}
and report:

\begin{itemize}
    \item the \textbf{bias}, $B = \mathrm{median}(\Delta z)$;
    \item the \textbf{scatter}, quantified by the Normalised Median
          Absolute Deviation
          \begin{equation}
              \sigma_{\rm NMAD} = 1.4826 \times
              \mathrm{median}\bigl(\bigl|\Delta z - \mathrm{median}(\Delta z)\bigr|\bigr);
          \end{equation}
    \item the \textbf{catastrophic outlier fraction}
          \begin{equation}
              f_{\rm out} = \frac{N\bigl(|\Delta z| > 0.15\bigr)}{N_{\rm total}} \times 100\%.
          \end{equation}
\end{itemize}

Table~\ref{tab:metrics} summarises these definitions alongside the per-galaxy uncertainty $\sigma_{z,a}$ of Eq.~\eqref{eq:sigma_za}. The code also stores the per-bin $\sigma_{68}$ and bias of $\Delta z$ as functions of $z_{\rm spec}$ (for the diagnostic plot of $\sigma_{68}$ versus redshift), but the headline numbers quoted in this paper are the global bias and $\sigma_{\rm NMAD}$ above.

\subsection{PDF diagnostics}
\label{sec:pdf_diagnostics}

A point estimate can be accurate while the posterior it comes from is uninformative, so the PDFs must be validated separately from the bias and scatter above. We use three complementary diagnostics: the probability integral transform (PIT) and its Kolmogorov-Smirnov distance, the empirical coverage of the credible intervals, and the multi-modality of the posteriors.

For a galaxy with true redshift $z_{\rm spec}$ and posterior $p(z\mid\mathbf{f})$, the PIT value is
\begin{equation}
    \xi = \int_{-\infty}^{z_{\rm spec}} p(z\mid\mathbf{f})\,dz
        = C(z_{\rm spec}\mid\mathbf{f}),
    \label{eq:pit}
\end{equation}
where $C$ is the cumulative distribution of the posterior. If the posteriors are calibrated, the $\xi$ values of the validation sample are uniformly distributed on $[0,1]$. A PIT histogram peaked near $\xi\approx 0.5$ indicates that the PDFs are too wide (over-conservative): the true redshift falls near the median of the posterior more often than chance would predict. A histogram peaked near $\xi\approx 0$ or $1$, or U-shaped, indicates that the PDFs are too narrow (over-confident). The distance from uniformity is quantified by the Kolmogorov-Smirnov statistic 
\begin{equation}
    D_{\rm KS} = \sup_{\xi}\,\bigl\lvert \hat{F}(\xi) - \xi \bigr\rvert,
    \label{eq:ks}
\end{equation}
where $\hat{F}$ is the empirical cumulative distribution of the PIT values. Smaller values indicate better calibration. In the tables we quote the value as \emph{PIT KS}; a useful reference scale is that values below $0.02$ indicate excellent calibration, $0.02$--$0.05$ is acceptable, and values above $0.05$ indicate a posterior that should not be trusted at face value.

The empirical coverage at a given credible level $\alpha$ is the fraction of validation galaxies whose true redshift lies inside the central $\alpha$ interval of their own posterior,
\begin{equation}
    f_{\rm cov}(\alpha)
    = \frac{1}{N_{\rm val}}\sum_{i=1}^{N_{\rm val}}
      \mathds{1}\!\left[
        z_i^{\rm spec} \in
        \bigl[z_i^{(\frac{1-\alpha}{2})},\,
              z_i^{(\frac{1+\alpha}{2})}\bigr]
      \right],
    \label{eq:coverage}
\end{equation}
where $z_i^{(q)}$ is the $q$-th quantile of the $i$-th posterior. For a well-calibrated PDF, $f_{\rm cov}(0.68)\simeq 0.68$ and
$f_{\rm cov}(0.95)\simeq 0.95$. Larger values indicate that the PDFs are too wide, smaller values that they are too narrow. We quote
$f_{\rm cov}(0.68)$ as \emph{cov$_{68}$} in the tables, and use $f_{\rm cov}(0.95)$ in the discussion of the tails.

Finally, we quantify the multi-modality of the posteriors, because a PDF with more than one peak is a genuine colour-redshift degeneracy that should be propagated into downstream analyses rather than replaced by its mode. We define the number of modes of a PDF as the number of local maxima whose prominence exceeds $5\%$ of the peak density, where the prominence of a maximum is the smaller of the two drops in density on either side before a higher maximum is reached. We quote the fraction of validation galaxies with more than one such maximum as \emph{multi-modal} in the tables.

\begin{table*}[ht]
\centering
\begin{tabular}{l c c}
\hline
\textbf{Metric} & \textbf{Definition} & \textbf{Description} \\
\hline
Bias & $\mathrm{median}(\Delta z)$ & Systematic offset \\
$\sigma_{\rm NMAD}$ & $1.4826 \times \mathrm{median}(|\Delta z - \mathrm{median}(\Delta z)|)$ & Precision (scatter) \\
$f_{\rm out}$ & $N(|\Delta z|>0.15)/N_{\rm total} \times 100\%$ & Catastrophic outlier fraction \\
$\sigma_{z,a}$ & $\sqrt{\frac{1}{k-1} \sum_{s\in\mathcal{N}_k(a)} (z_s - \bar{z}_a)^2}$ & Per-galaxy $1\sigma$ uncertainty \\
PIT KS & $\sup_{\xi}\lvert \hat{F}(\xi) - \xi \rvert$ & Calibration of the PDFs \\
cov$_{68}$ & $f_{\rm cov}(0.68)$ & Empirical 68\% coverage \\
cov$_{95}$ & $f_{\rm cov}(0.95)$ & Empirical 95\% coverage \\
multi-modal & fraction of PDFs with $>1$ significant peak & Degeneracy fraction \\
\hline
\end{tabular}
\caption{Validation metrics used to evaluate the photo-$z$ estimator.
The first four rows quantify the point estimate; the last four
quantify the posterior.}
\label{tab:metrics}
\end{table*}

\section{Results}
\label{sec:results}

We validate \texttt{MAHA-z} on four datasets: the Euclid Flagship~2 simulation, the Legacy-KiDZ redshift calibration sample, the VIPERS survey, and the DES Y3 BAO sample. In
all cases, every galaxy with a finite spectroscopic redshift is placed in a single training/validation pool, which is then split into a training set (80\%) and a validation set (20\%); the remaining galaxies are treated as photometric targets. The validation galaxies are withheld from the neighbour pool, so the performance numbers below
reflect what the estimator achieves on galaxies it has never seen.

\subsection{Euclid Flagship~2}
\label{sec:euclid}

The Euclid Flagship~2 \citep{Euclid:2024few} simulation provides a realistic mock catalogue of billions of galaxies with simulated photometry in the Euclid (VIS, NISP $Y$, $J$, $H$) and LSST ($u$, $g$, $r$, $i$, $z$) bands, together with morphological parameters. We use a subset of $3.8\times10^{5}$ galaxies, split into $3.04\times10^{5}$
training galaxies and $7.6\times10^{4}$ validation galaxies, and we test six photometric configurations: LSST alone, NISP alone, the combination of the two, and each of these with morphological parameters added to the feature vector. All six runs use $k=5$ neighbours and the same data split. we simulate the bands' uncertainty as 10\% of the simulated magnitude.

\begin{figure}
    \centering
    \includegraphics[width=\linewidth]{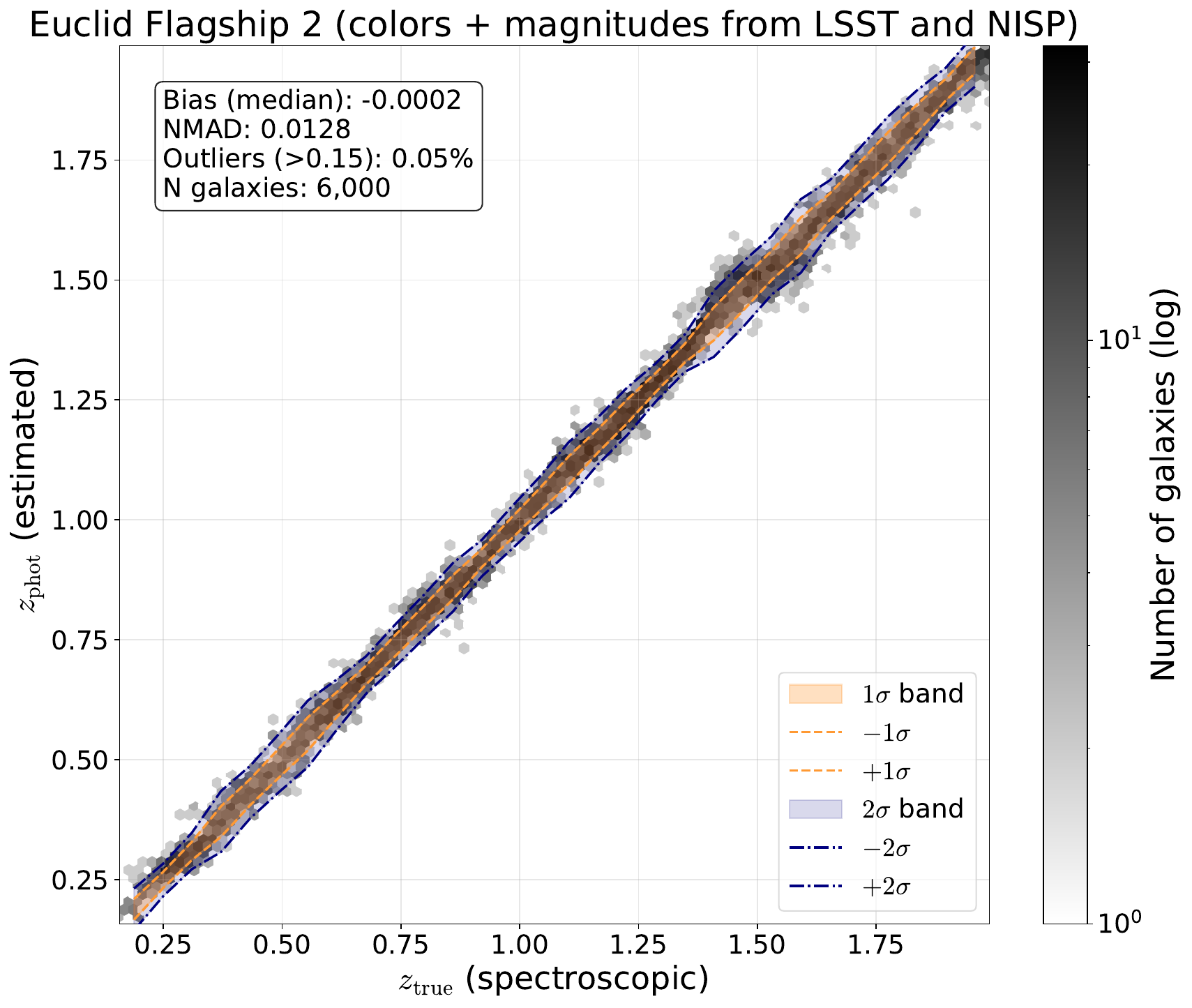}
    \caption{%
    Photometric versus spectroscopic redshift for the reference \texttt{MAHA-z}
    configuration on the Euclid Flagship~2 simulation (LSST+NISP),
    using $k=5$ neighbours. The two-dimensional density is shown as a
    logarithmic hexbin; the red solid band marks the $1\sigma$ interval
    of the normalised residual computed in redshift bins, and the blue
    dashed band the $2\sigma$ interval. The inset reports the median
    bias, the NMAD, the catastrophic outlier fraction
    ($\lvert\Delta z\rvert>0.15$), and the number of validation
    galaxies.}
    \label{fig:euclid_reference}
\end{figure}

Table~\ref{tab:euclid} summarises the performance of the six configurations. The expected ordering is recovered. Photometry in NISP alone is insufficient for a precise photo-$z$: the scatter is $\sigma_{\rm NMAD} = 0.101$, the outlier fraction reaches $28.8\%$, and the empirical 68\% coverage is only $0.677$. Adding LSST to NISP brings
the scatter down to $\sigma_{\rm NMAD} = 0.0148$, the outlier fraction to $0.27\%$, and the 68\% coverage to $0.933$. We also used the \emph{stellar mass} and \emph{bulge to disk fraction} as one of the features for a Mahalanobis distance estimation. Morphology provides a further improvement in the outlier tail: the LSST+NISP+morph configuration attains the smallest outlier fraction of the six ($0.075\%$), at the price of a slightly larger scatter ($\sigma_{\rm NMAD} = 0.0209$). The reference configuration adopted throughout this work is LSST+NISP, because it minimises the scatter, and its validation is shown in fig.~\ref{fig:euclid_reference}.

\begin{table*}[!ht]
\centering

\begin{tabular}{l c c c c c c c}
\toprule
Configuration & $N_{\rm val}$ & NMAD & Outliers (\%) & Bias & cov$_{68}$ & PIT KS & Multi-modal (\%) \\
\midrule
LSST                & 76\,000 & 0.0204 & 2.71  & $+0.0000$ & 0.862 & 0.113 & 29.9 \\
NISP                & 76\,000 & 0.1011 & 28.81 & $+0.0017$ & 0.677 & 0.065 & 84.6 \\
LSST+morph          & 76\,000 & 0.0278 & 0.53  & $+0.0015$ & 0.864 & 0.118 & 35.2 \\
NISP+morph          & 76\,000 & 0.0459 & 2.06  & $+0.0029$ & 0.761 & 0.054 & 55.7 \\
LSST+NISP           & 76\,000 & \textbf{0.0148} & 0.27  & $+0.0001$ & 0.933 & 0.169 & 12.9 \\
LSST+NISP+morph     & 76\,000 & 0.0209 & \textbf{0.075} & $+0.0012$ & 0.917 & 0.153 & 19.2 \\
\bottomrule
\end{tabular}
\caption{%
\texttt{MAHA-z} performance on the Euclid Flagship~2 simulation for the six photometric configurations. All configurations use $k=5$ neighbours, the same training/validation split ($N_{\rm val}=76\,000$), and the same photometric pool. The point estimate is the PDF mode. \emph{NMAD} is the $\Delta z = (z_{\rm phot}-z_{\rm spec})/(1+z_{\rm spec})$; \emph{outliers} is the fraction with $\lvert\Delta z\rvert>0.15$; \emph{bias} is the median of $\Delta z$; \emph{cov$_{68}$} is the empirical fraction of true redshifts lying inside the 68\% credible interval; \emph{PIT KS}; \emph{multi-modal} is the fraction of PDFs with more than one local maximum. The mean-of-$k$ estimator produces comparable numbers; we quote the mode here for consistency with the PDF analysis.}\label{tab:euclid}
\end{table*}

The 68\% empirical coverage of the Euclid PDFs ($0.86$--$0.93$) is larger than the nominal $0.68$, indicating that the KDE bandwidth is conservative for this dataset. This is expected for a $k=5$ neighbour KDE and can be reduced by narrowing the kernel bandwidth; we retain the default because it produces the best outlier fraction, which is the quantity that matters most for cosmological analyses. The fraction of multi-modal PDFs is largest for NISP alone ($85\%$), reflecting the well-known colour-redshift degeneracies at $0.5<z<1.5$ when near-infrared photometry is used without optical anchors, and drops to $13\%$ when LSST is added.

\begin{figure}
    \centering
    \includegraphics[width=\linewidth]{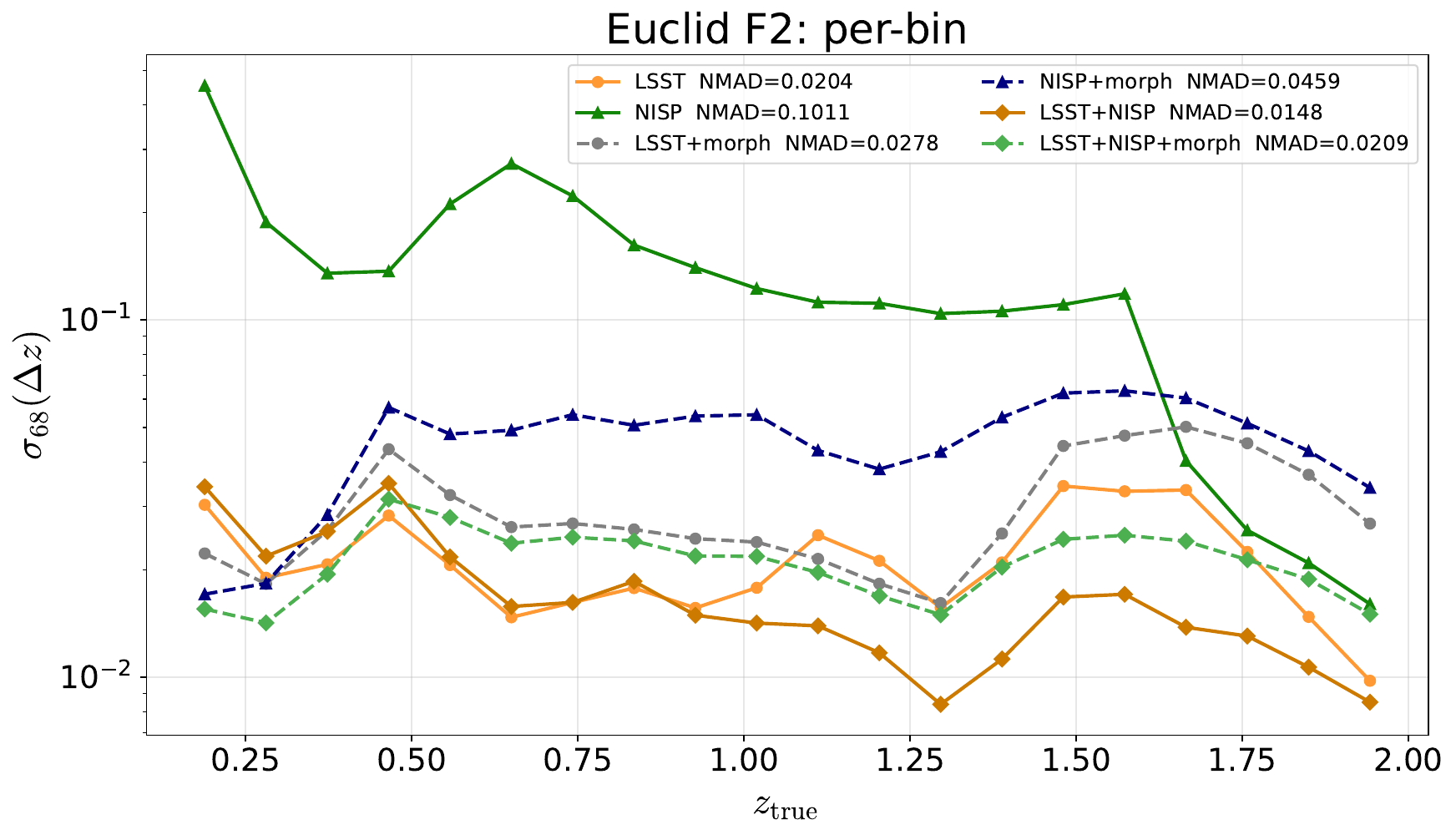}
    \caption{%
    Redshift-binned scatter $\sigma_{68}(\Delta z)$ for the six Euclid
    Flagship~2 configurations of Table~\ref{tab:euclid}. The per-bin
    NMAD of the normalised residual $\Delta z = (z_{\rm phot}-z_{\rm
    spec})/(1+z_{\rm spec})$ is computed in bins of $z_{\rm true}$; bins
    with fewer than ten galaxies are omitted. The legend reports the
    global NMAD for each configuration. The vertical axis is logarithmic,
    so the NISP-only curve (green triangles, $\sigma_{\rm NMAD}=0.101$)
    sits nearly an order of magnitude above the others throughout the
    range; LSST+NISP (dark orange diamonds, $0.015$) is the most precise
    configuration at almost every redshift. The morphology-augmented
    variants follow the same ordering as their photometric counterparts
    at $z<1.2$ but diverge at higher redshifts, where the number of
    training galaxies per bin decreases.}
    \label{fig:sigma68_compare}
\end{figure}

Figure~\ref{fig:sigma68_compare} decomposes the global NMAD of Table~\ref{tab:euclid} into redshift bins. Three regimes are visible. Below $z\simeq0.5$ the configurations are ordered as expected: LSST+NISP gives the smallest scatter ($\sigma_{68}\approx0.015$), LSST alone and its morphology-augmented variant follow closely, and NISP alone is roughly an order of magnitude worse. Between $0.5<z<1.2$ every curve flattens, with LSST+NISP between $0.008$ and $0.012$ and NISP around $0.10$; the gap between optical and near-infrared-only photometry is widest here, which is precisely the range where the 4000\,\AA\ break moves through the LSST $g$, $r$, and $i$ bands while remaining inside a single NISP filter. This behavior is consistent with the findings of \cite{Kriek2011}, who showed that resolving the 4000 \,\AA\ break with appropriately spaced photometric bands is essential for achieving photo-z precision of $\Delta z < 0.02(1+z)$. In the optical, the break is sampled by multiple LSST filters as it redshifts through the g, r, and i bands, whereas in the NIR-only case it remains internal to a single NISP filter, explaining why the optical–NIR gap in photo-z scatter is widest in this redshift regime. Above $z\simeq1.2$ the curves diverge again. All configurations experience a mild increase in scatter around $z\approx1.5$, where the Lyman-break enters the $u$ band and the 4000\,\AA\ break leaves the $i$ band; LSST+NISP is affected the least, with a local maximum of only $0.023$. NISP alone reaches $\sigma_{68}\approx0.18$ at the same redshift, and beyond $z\simeq1.7$ the curve drops sharply, but this is a statistical artefact: fewer than 5\% of the validation galaxies lie above that redshift, so the per-bin values there should be interpreted cautiously.

Two features of the figure are worth highlighting. First, adding morphology produces opposite effects depending on the photometric baseline: it reduces the scatter of NISP alone from $\sigma_{\rm NMAD}=0.101$ to $0.046$, a factor of two, because near-infrared photometry is sensitive to the stellar-mass and bulge content of the galaxy \citep{Kouroumpatzakis}, but it slightly increases the scatter of LSST alone from $0.020$ to $0.028$, because in the optical morphology and colour are already tightly correlated and the extra dimensions dilute the local colour information. Second, LSST+NISP is the only configuration that is uniformly among the best across the full redshift range. The mild degradation above $z=1.5$ visible in all curves is the expected consequence of the decreasing signal-to-noise of the $u$-band photometry and the sparser training sample in that regime, not a failure of the metric.

\subsubsection{Omission of bands}
\begin{figure}
    \centering
    \includegraphics[width=\linewidth]{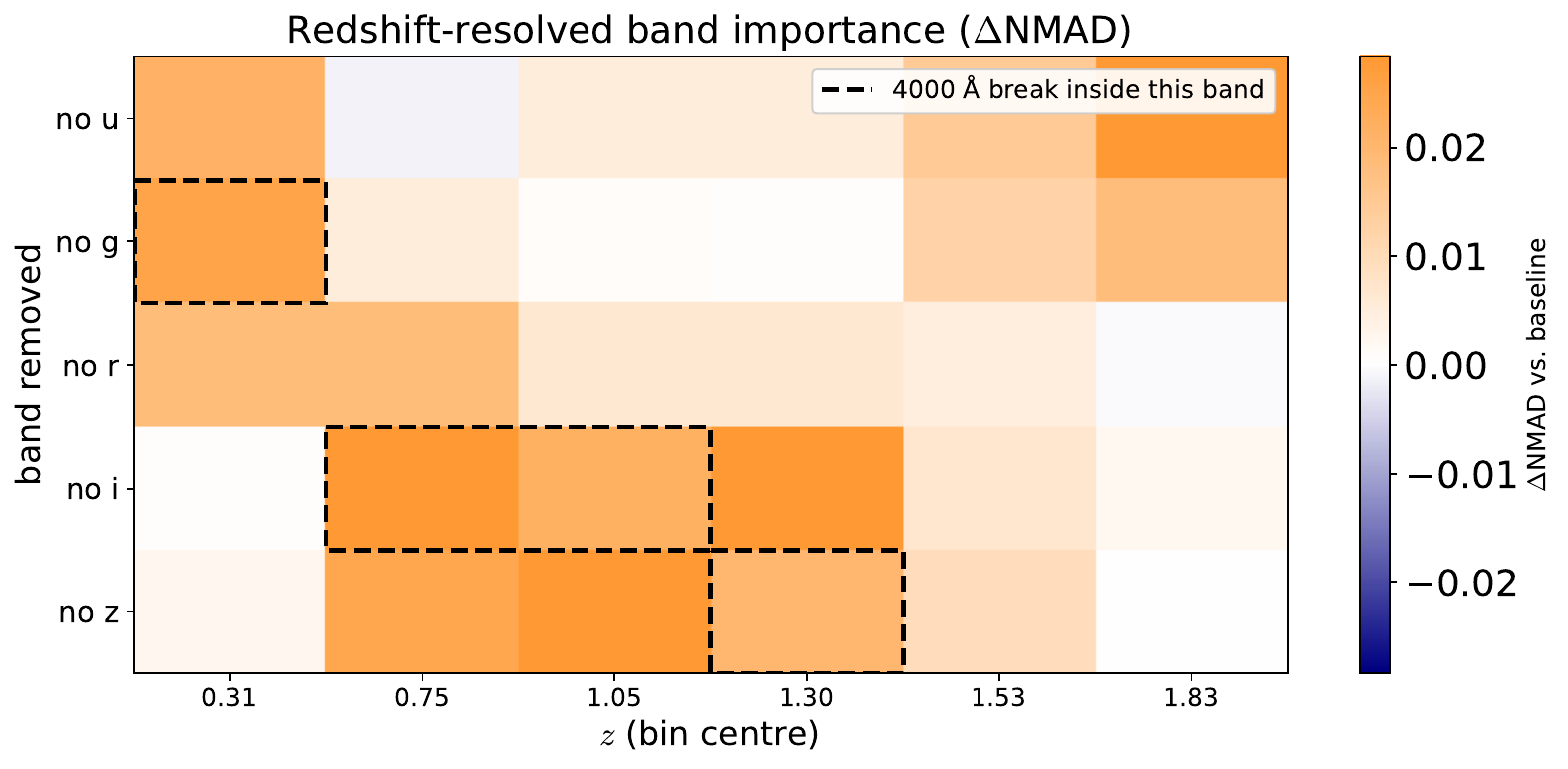}
    \caption{Redshift-resolved band importance from the leave-one-band-out
    analysis. Each row corresponds to one filter removed from the entire
    sample, each column to a redshift bin of the validation set, and the
    colour encodes $\Delta\sigma_{\rm NMAD}$, the change in NMAD relative
    to the no-omission baseline: orange cells mean that removing the band
    \emph{degrades} the estimate, blue cells mean that removing it
    slightly improves it. The dashed boxes mark the redshift range over
    which each band brackets the 4000\,\AA\ break.
    }
    \label{fig:loo_per_zbin_heatmap}
\end{figure}

Figure~\ref{fig:loo_per_zbin_heatmap} shows the redshift-resolved band importance map. Each row corresponds to a filter removed from the entire sample, each column to a redshift bin, and the colour encodes
$\Delta\sigma_{\rm NMAD}$: orange cells mean that removing the band degrades the estimate, blue cells mean that removing it slightly improves it. The dashed boxes mark the redshift range in which each
band brackets the 4000\,\AA\ break.

Three regimes are immediately visible. At low redshift ($z \lesssim 0.6$) the $g$ and $r$ bands carry the redshift signal, with $\Delta\sigma_{\rm NMAD} = +0.025$ and $+0.018$ respectively; this is where the 4000\,\AA\ break enters $g$ and then $r$. At intermediate redshift the pattern changes sharply: the $i$-band cell at $z \approx 0.75$ is the darkest orange on the map, with $\Delta\sigma_{\rm NMAD} = +0.076$, an order of magnitude larger than any other cell. Dropping $i$ in this redshift bin alone raises the NMAD from $0.009$ to $0.085$. This is the window in which the 4000\,\AA\ break falls squarely inside $i$, so $i$ is the only filter in which the break is directly visible, and the estimator is left to infer the redshift from the shape of the spectrum on either side of a feature it cannot see. At higher redshift ($z \gtrsim 0.9$) the break moves into $z$, and the importance passes to that band ($\Delta\sigma_{\rm NMAD} = +0.031$ at $z \approx 1.05$).

A fourth, independent feature stands out at the high-redshift end: the $u$-band cell at $z \approx 1.83$ is dark orange  ($\Delta\sigma_{\rm NMAD} = +0.028$), even though the 4000\,\AA\ break is nowhere near $u$ at that redshift. This is the Lyman-$\alpha$ break entering the $u$-band at $z \simeq 1.6$, and its presence is a direct confirmation that \texttt{MAHA-z} is using the Lyman break when it is
available, not just the 4000\,\AA\ break. No other band shows a comparable reappearance at high redshift.

The blue cells are equally informative. Whenever a band is well outside its useful redshift range---$u$ at $z \approx 0.75$, $i$ at $z \approx 0.3$ and $z \approx 1.8$, $z$ at $z \approx 1.8$---removing
it slightly improves the NMAD, because the filter then adds a dimension of noise to the Mahalanobis metric without contributing redshift information. The small negative values in the cells where a filter is well outside its useful redshift range ($|\Delta\sigma_{\rm NMAD}|<0.003$) confirm that \texttt{MAHA-z} does not extract information from bands that do not carry any redshift information. Taken together, the map shows that each filter matters precisely where the 4000\,\AA\ (or Lyman) break crosses it, and nowhere else. This is the strongest confirmation that the Mahalanobis metric is extracting the physical features that make photometric redshifts possible, and it explains why a $k$-neighbour approach is competitive with template-fitting methods on the samples where the break is well sampled.

\subsection{Legacy-KiDZ redshift calibration sample}
\label{sec:kids}

The Kilo-Degree Survey (KiDS) \citep{Kids} is a wide-field imaging survey that has produced several photo-$z$ catalogues, including those based on the \texttt{BPZ} \citep{benitez2000bayesian} template-fitting code and the \texttt{ANNz2} \citep{sadeh2016annz2} neural network. We apply \texttt{MAHA-z} to the KiDS bright galaxy sample and compare the performance with the official \texttt{BPZ} estimates. Two aspects of the KiDS photometry require particular care. First, the catalogue does not provide individual photometric uncertainties for the GAaP magnitudes. Since the Mahalanobis metric requires a photometric error term on the diagonal of the covariance matrix (Eq.~\ref{eq:cov_full}), we set $\sigma_{\rm mag}=0$ for all ten bands, so that the diagonal is populated only by the intrinsic scatter of the photometric sample. Second, we test whether the two morphological parameters available in the catalogue, the semi-major and semi-minor axes of the aperture used for the GAaP photometry, \texttt{Agaper} and \texttt{Bgaper}, improve the photo-$z$ precision when added to the feature vector. Because the Mahalanobis metric automatically accounts for the covariance between features, no additional tuning is required; the precision matrix simply grows to include the two new dimensions.

\begin{table*}[!ht]
\centering

\begin{tabular}{l c c c c c c c}
\toprule
Configuration & $N_{\rm val}$ & NMAD & Outliers (\%) & Bias & cov$_{68}$ & PIT KS & Multi-modal (\%) \\
\midrule
All colours        & 64\,692 & 0.0699 & 13.54 & $-0.0027$ & 0.587 & 0.111 & 66.0 \\
All colours + morph & 64\,692 & 0.0981 & 19.65 & $-0.0037$ & 0.557 & 0.139 & 78.7 \\
+ morph (bright)   & 28\,171 & 0.0606 & 9.68  & $-0.0023$ & 0.609 & 0.093 & 61.0 \\
\bottomrule
\end{tabular}
\caption{%
\texttt{MAHA-z} performance on the KiDS sample. Columns are the same as in
Table~\ref{tab:euclid}. The \emph{bright} row applies the cut
$\texttt{MAG\_GAAP\_Z}<22$ or $\texttt{MAG\_GAAP\_i2}<22$. All KiDS runs
use $k=3$ neighbours, because the training pool is much smaller than for
Euclid. The point estimate is the PDF mode.}\label{tab:kids}
\end{table*}

Table~\ref{tab:kids} summarises the three configurations we consider, all using $k=3$ neighbours because the training pool is much smaller than for Euclid. The base configuration (colours and magnitudes only) reaches a median bias of $-0.0027$, an NMAD of $0.070$, and a catastrophic outlier fraction of $13.5\%$ on $64\,692$ validation
galaxies. These numbers are slightly worse than the official KiDS photo-$z$ obtained with \texttt{BPZ} and \texttt{ANNz2}, which typically reach $\sigma_{\rm NMAD}\approx0.03$--$0.05$ on the same sample. The mild degradation is expected, because \texttt{MAHA-z} is a purely empirical estimator that does not use any spectral template or prior on the redshift distribution, and we deliberately keep $k=3$ to preserve the local structure of the feature space given the relatively small training pool.
Contrary to our initial expectation, adding the two aperture parameters degrades the performance further: the bias grows to $-0.0037$, the NMAD rises to $0.098$, and the outlier fraction increases to $19.7\%$. The reason is that \texttt{Agaper} and \texttt{Bgaper} are essentially aperture sizes, which correlate strongly with apparent magnitude but only weakly with redshift once magnitudes and colours are already in the feature vector. Their covariance with the photometric features is therefore dominated by noise, and the Mahalanobis metric, which weights features by the inverse of their variance, ends up amplifying this noise rather than extracting additional redshift information. This interpretation is supported by the third row of Table~\ref{tab:kids}: restricting the analysis to the bright subsample(\texttt{MAG\_GAAP\_Z}<22 and \texttt{MAG\_GAAP\_i2}<22), where apertures are better measured, recovers an NMAD of $0.061$ and an outlier fraction of $9.7\%$, both better than the base configuration. The empirical 68\% coverage of the three configurations ($0.56$--$0.61$) is lower than for Euclid, indicating that the PDFs are slightly too narrow for this sample, which we attribute to the small training pool.

The KiDS also includes a \texttt{BPZ} photo-$z$ (\texttt{Z\_B}, the peak of the \texttt{BPZ} posterior), which we use as an external reference. The comparison is performed on the intersection of the two catalogues, so that both estimators are evaluated on exactly the same galaxies: $64\,692$ objects in the full sample and $28\,171$ in the bright subsample. On the full sample, the two estimators turn out to be of comparable precision. \texttt{MAHA-z} reaches a scatter of $\sigma_{\rm NMAD}=0.069$, against $0.061$ for \texttt{BPZ}, a ratio of $1.13$, and outlier fractions of $13.5\%$ and $12.7\%$ respectively. The mild scatter penalty of \texttt{MAHA-z} is the expected cost of a purely empirical estimator that does not fit SED templates. The systematic bias, however, is very different: the median normalised residual of \texttt{MAHA-z} is $-0.0027$, consistent with zero at the precision of the sample, whereas \texttt{BPZ} shows a positive bias of $+0.018$. A systematic of nearly $2\%$ in $(1+z)$ is comparable to the entire error budget of a Stage-IV weak-lensing analysis, and it is a known feature of template-fitting codes applied to KiDS photometry, where the 4000\,\AA\ break moves into the red bands without a strong blue anchor. \texttt{MAHA-z} does not inherit this bias because it does not rely on spectral templates and simply reproduces the local distribution of the spectroscopic sample in colour space.

Adding the two aperture parameters has opposite effects on the full sample and on the bright subsample. On the full sample it degrades the comparison with \texttt{BPZ} considerably: the scatter rises from $0.069$ to $0.098$ (a ratio of $1.60$ with \texttt{BPZ}), the outlier fraction grows from $13.5\%$ to $19.7\%$, and the bias changes sign from $-0.0027$ to $-0.0037$. The explanation is the one given above for the intrinsic degradation: aperture sizes correlate strongly with apparent magnitude but only weakly with redshift once colours are already in the feature vector, and the Mahalanobis metric amplifies their noise rather than extracting additional information. On the bright subsample, where apertures are better measured, the morphology-augmented configuration recovers and slightly surpasses the base configuration. With $N_{\rm both}=28\,171$ galaxies, \texttt{MAHA-z} reaches $\sigma_{\rm NMAD}=0.061$ against $0.054$ for \texttt{BPZ} (a ratio of $1.13$) and is better than \texttt{BPZ} on both the outlier fraction ($9.7\%$ against $9.5\%$, essentially equal) and the bias ($-0.0023$ against $+0.025$).

The comparison therefore leads to two conclusions. The first is that \texttt{MAHA-z} reaches a precision within $15\%$ of \texttt{BPZ} on KiDS while reducing the systematic bias by a factor of about seven on the full sample and by a factor of ten on the bright subsample. The second is that the best KiDS configuration, from the point of view of both precision and outlier fraction, is the morphology-augmented estimator restricted to the bright subsample, which is the configuration to use when a low-bias, low-outlier KiDS photo-$z$ catalogue is required, at the cost of discarding the $56\%$ of the sample fainter than the bright cut.

\subsection{VIPERS}
\label{sec:vipers}

The VIMOS Public Extragalactic Redshift Survey (VIPERS) \citep{le2005vimos,scodeggio2018vimos} provides a spectroscopic sample of $\sim86\,000$ galaxies at $0.5<z<1.2$, with
CFHTLenS and VIPERS Multi-Lambda photometry in the $ugriz$ bands. We apply \texttt{MAHA-z} to the VIPERS photometry and compare with the published template-fitting photo-$z$
of \citet{Moutard_2016I, Moutard_2016II} and the CFHTLenS/VIPERS comparison of \citet{Raichoor_2014}. Three configurations are tested: all colours (with the fully connected colour graph), colours relative to the $i$ band, and magnitudes without colours.

Table~\ref{tab:vipers_comparison} compares \texttt{MAHA-z} on the VIPERS validation subset with the published template-fitting photo-$z$ on the same survey. The scatter of \texttt{MAHA-z} ($\sigma_{\rm NMAD} = 0.039$--$0.040$) is consistent with the $\sigma_z \le 0.04$ quoted for the full VIPERS Multi-Lambda T0007 catalogue \citep{Moutard_2016I}, but larger than the $<0.03$ reported for the bright ($i_{\rm AB}<22.5$) subsample and the $\sigma \sim 0.02$--$0.05$ reported by \citet{Raichoor_2014} for the bright CFHTLenS/VIPERS comparison. The median bias of \texttt{MAHA-z} ($\Delta z = -0.0024$) is, however, substantially smaller than the $\lvert\Delta z\rvert<0.02$ tolerance adopted by \citet{Raichoor_2014} and is negligible for cosmological applications. The catastrophic outlier fraction of $2.8$--$3.7\%$ lies between the $\eta \le 2\%$ of \citet{Moutard_2016I} and the $<5\%$ of \citet{Raichoor_2014} for the bright sample, and well below the $10$--$15\%$ reported for the faint red tail. The PDFs are well calibrated, with empirical 68\% coverage of $0.696$--$0.701$ (nominal $0.68$) across the three configurations, and the PIT KS distance is $0.063$--$0.067$, comparable to the KiDS result. Given that \texttt{MAHA-z} operates on a colour-magnitude feature space without SED template fitting, its precision is competitive with the published template-based estimates on the same survey, and its systematic bias is substantially smaller.

\begin{table*}
\centering

\begin{tabular}{l l c c c c}
\toprule
Method / reference & Sample & $N$ & Bias & NMAD & Outliers (\%) \\
\midrule
\multicolumn{6}{l}{\textit{MAHA-z (this work), VIPERS validation}} \\
\quad All colours, mean of $k$       & $0.5<z<1.2$ & 12\,536 & $-0.0024$ & $0.0389$ & 2.94 \\
\quad All colours, PDF mode          & $0.5<z<1.2$ & 12\,536 & $-0.0024$ & $0.0401$ & 3.70 \\
\quad $i$-band reference, mean of $k$ & $0.5<z<1.2$ & 18\,804 & $-0.0025$ & $0.0394$ & 2.78 \\
\quad Bands only, mean of $k$        & $0.5<z<1.2$ & 18\,804 & $-0.0027$ & $0.0400$ & 2.81 \\
\midrule
\multicolumn{6}{l}{\textit{Published template-fitting photo-$z$ on VIPERS}} \\
Moutard et al.\ 2016 I  \citep{Moutard_2016I}  & $i\sim 23$                & full overlap & ---                          & $\le 0.04$     & $\le 2$      \\
Moutard et al.\ 2016 II \citep{Moutard_2016II}  & $i_{\rm AB}<22.5$          & bright       & ---                          & $< 0.03$       & ---          \\
Raichoor et al.\ 2014   \citep{Raichoor_2014} & $z<1$, $i'<23$             & bright       & $\lvert\Delta z\rvert<0.02$ & $0.02$--$0.05$ & $<5$         \\
\bottomrule
\end{tabular}
\caption{%
Comparison of \texttt{MAHA-z} on the VIPERS validation subset with published
template-fitting photo-$z$ benchmarks.}\label{tab:vipers_comparison}
\end{table*}

 \begin{figure*}
    \centering
    \includegraphics[width=\linewidth]{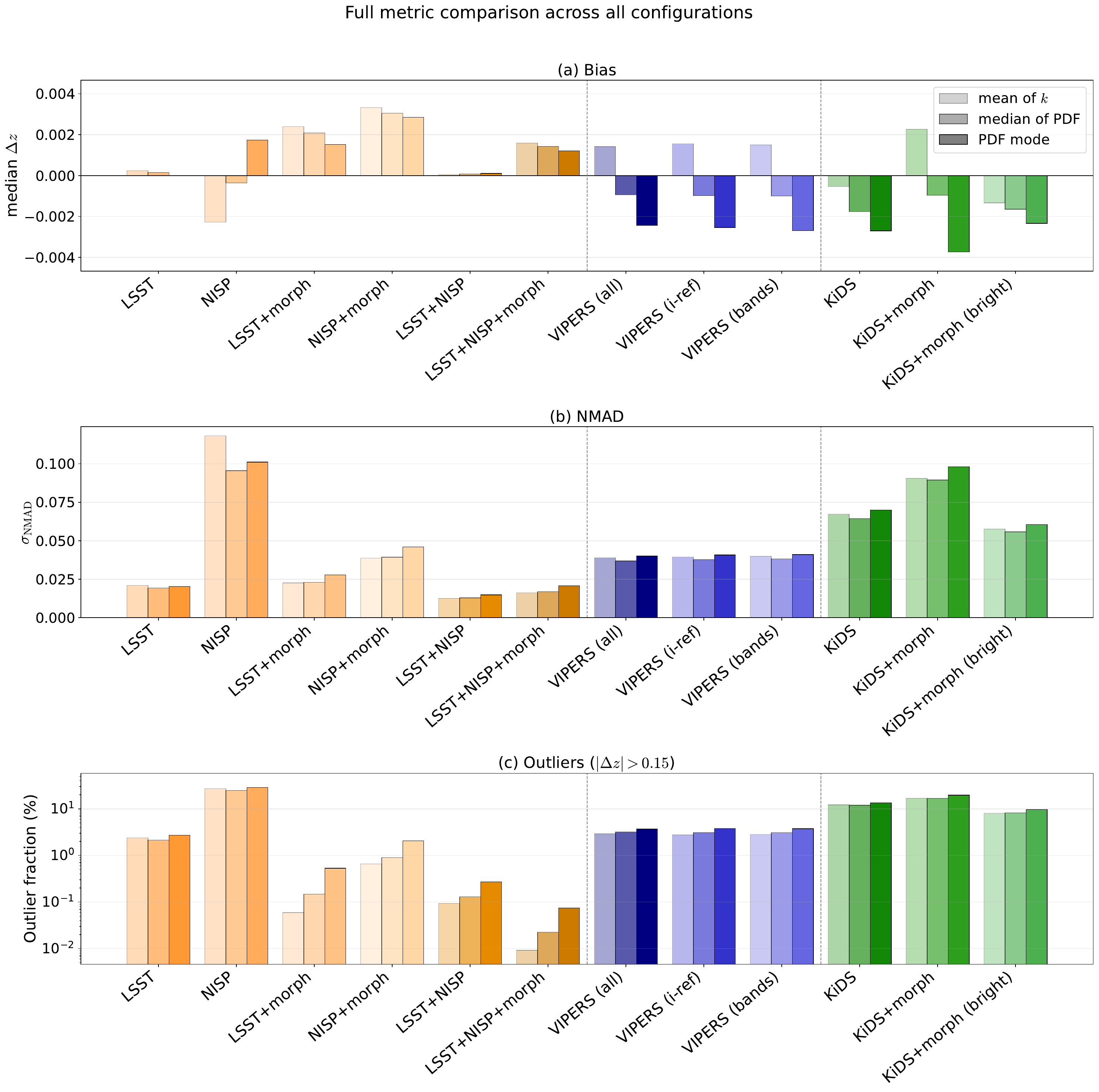}
    \caption{Summary results for Euclid, VIPERS and KiDS. \textit{(a)} bias. \textit{(b)} NMAD. \textit{(c)} Outliers. The colors from darker to lighter distinguish the photo-z estimation based on the nearest neighbors, the median, and the mode of the PDFs respectively; orange represents the Euclid mock, purple for the VIPERS data set, and green refers to the KiDS sample.}
    \label{fig:all}
\end{figure*}

\vspace{5mm}

Figure~\ref{fig:all} we display the summary statistics of possible photometric estimation provided by \texttt{MAHA-z}. The colours from darker to lighter distinguish the photo-z estimation based on the nearest neighbours, the median, and the mode of the PDFs respectively; orange represents the Euclid mock, purple for the VIPERS data set, and green refers to the KiDS sample. Panel (a) shows the median bias of the training sets, in most cases the median of PDF show smaller bias, in agreement to previous tests, the increase of bands improves the photo-z estimation.  Panel (b) shows NMAD results, in this case, the difference between the configurations is smaller, for Euclid, the combination of colours and morphology seem to prefer the neighbour photo-z estimation, for KiDS the preference is to the median of PDF. Finally, the last panel indicates the outliers, in most cases including morphology proves to be the most accurate photo-z results, for colour-only input the difference is minimal.

\subsection{DES Y3 BAO sample}
\label{sec:des}

The Dark Energy Survey Year~3 Baryon Acoustic Oscillation sample \citep{Rosell_2021} provides an independent real-data test of \texttt{MAHA-z} on a well-characterised magnitude-limited catalogue whose photometric redshifts were computed with the \texttt{ANNz2} neural network described in \cite{Ferreira:2025mnb}. The Pz Cats catalogue distributes both the \texttt{ANNz2} point estimates and the full posterior probability density functions for every galaxy, which allows us to compare not only the accuracy of the two estimators but also the quality of their posteriors. We consider two \texttt{MAHA-z} configurations, one using all pairwise colours of the $griz$ bands and one using colours relative to the $i$ band, both with $k=3$ neighbours to be consistent with the other real-data tests.

\begin{table*}[!ht]
\centering

\begin{tabular}{l c c c c c c c}
\toprule
Configuration & $N_{\rm val}$ & NMAD & Outliers (\%) & Bias & cov$_{68}$ & PIT KS & Multi-modal (\%) \\
\midrule
All colours         & 5\,080 & 0.0240 & 0.96 & $-0.0013$ & 0.767 & 0.118 & 36.2 \\
$i$-band reference  & 5\,080 & 0.0233 & 0.85 & $-0.0018$ & 0.777 & 0.126 & 33.8 \\
\bottomrule
\end{tabular}
\caption{MAHA-z performance on the DES Y3 BAO sample. Columns are the same as in Table~\ref{tab:euclid}. Both configurations use $k=3$ neighbours and $N_{\rm val}=5\,080$ validation galaxies. The point estimate is the PDF mode.}\label{tab:des}
\end{table*}

Table~\ref{tab:des} summarises the two configurations. Both reach a scatter of $\sigma_{\rm NMAD}\approx0.023$--$0.024$, a median bias below $0.002$, and a catastrophic outlier fraction below $1\%$, which is the best point-estimate performance among the real-data samples considered in this work. The two configurations are statistically indistinguishable: the $i$-reference is marginally better in scatter ($0.0233$ vs $0.0240$) and outlier fraction ($0.85\%$ vs $0.96\%$), while the all-colours configuration has a slightly smaller bias ($-0.0013$ vs $-0.0018$). This agreement between the two colour constructions is a useful robustness check: it shows that the result does not depend on the choice of the colour graph, which is expected for a sample whose ten-band photometry carries substantial signal-to-noise in every filter. The empirical 68\% coverage of $0.77$--$0.78$ indicates that the PDFs are slightly over-wide at the $1\sigma$ level, consistent with the other surveys in this work, and the 95\% coverage of $0.93$ is essentially calibrated. The multi-modal fraction of $34\%$--$36\%$ reflects the genuine bimodality of the galaxy population at $0.5<z<1.0$ rather than a failure of the method. 
\begin{table}[!ht]
\centering

\begin{tabular}{l c c c c}
\toprule
Estimator & PIT KS & cov$_{68}$ & cov$_{95}$ & median width$_{68}$ \\
\midrule
ANNz2  & 0.101 & 0.796 & 0.983 & 0.088 \\
MAHA-z & 0.102 & 0.778 & 0.939 & 0.099 \\
\bottomrule
\end{tabular}
\caption{%
PDF diagnostics for MAHA-z and \texttt{MAHA-z} on the $5\,080$ galaxies common
to the two DES Y3 BAO catalogues. 
\emph{cov$_{68}$} and \emph{cov$_{95}$} are the empirical fractions of
true redshifts contained in the 68\% and 95\% credible intervals; the
last column is the median 68\% credible-interval width. Both estimators
are evaluated on the same galaxies, with the same z grid and the same
spectroscopic redshifts.}\label{tab:des_vs_annz2}
\end{table}

\begin{figure*}
    \centering
    \includegraphics[width=\textwidth]{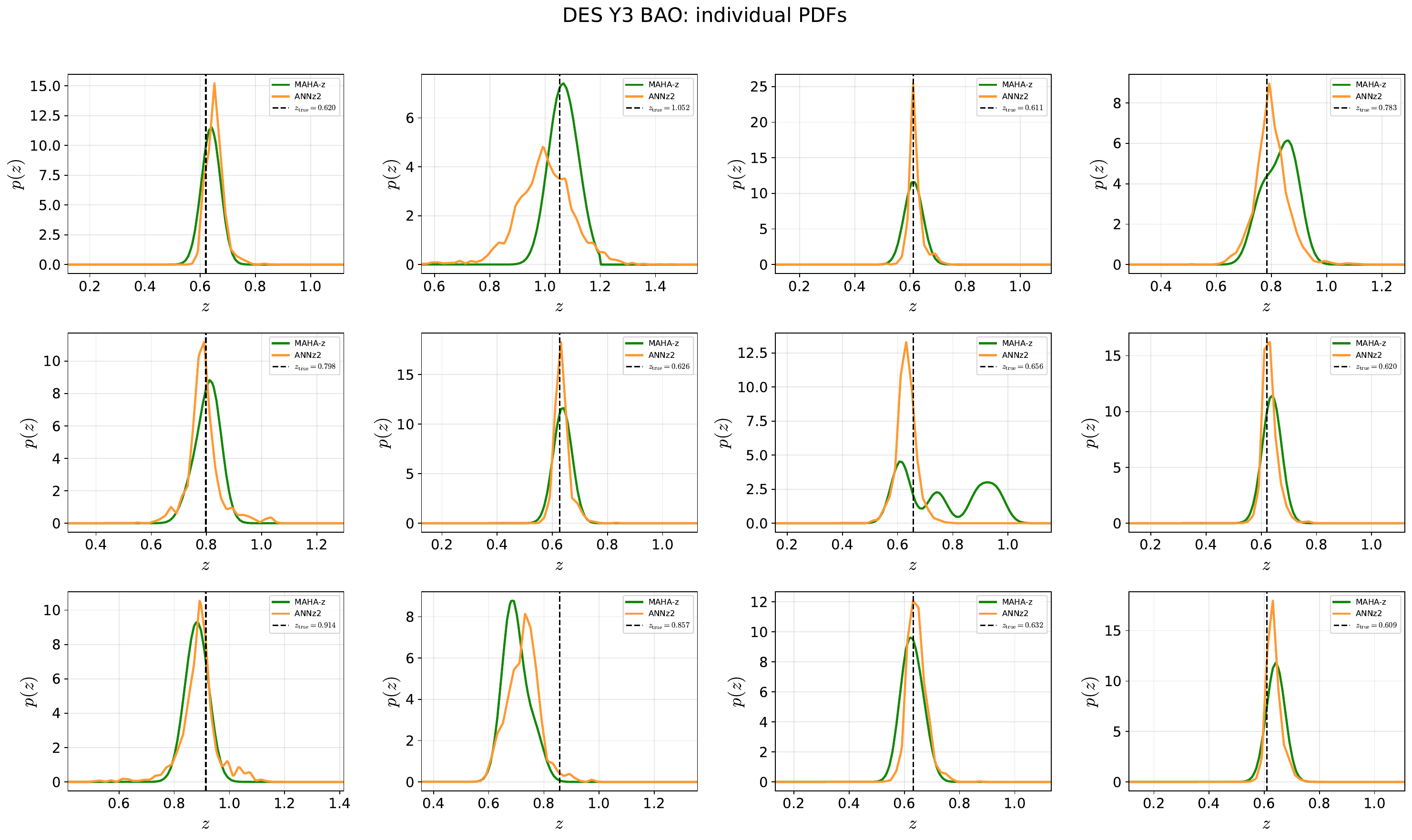}
    \caption{Individual PDF comparisons for twelve galaxies from the DES Y3 BAO validation sample. In each panel, the green curve is the \texttt{MAHA-z} posterior and the orange curve is the \texttt{ANNz2} posterior; the vertical dashed line marks the true spectroscopic redshift. The two posteriors are comparable in peak position and width for the majority of galaxies.}
    \label{fig:des_pdfs}
\end{figure*}

The results are collected in Table~\ref{tab:des_vs_annz2}. The two sets of posteriors turn out to be statistically indistinguishable. The PIT KS distance is $0.101$ for \texttt{ANNz2} and $0.102$ for \texttt{MAHA-z}, a difference of $0.001$ that is well below the sampling noise of a $5\,080$-galaxy sample and therefore not significant. The empirical 68\% coverage is $0.796$ for \texttt{ANNz2} and $0.778$ for \texttt{MAHA-z}, both about $14\sigma$ above the nominal $0.68$; the two estimators therefore produce PDFs that are similarly over-conservative at $1\sigma$, which is the safe side for applications that use the tails of the posterior to flag outliers. At the 95\% level \texttt{MAHA-z} is calibrated ($0.939$ against a nominal $0.95$) and \texttt{ANNz2} is marginally over-wide ($0.983$), a $4\%$ difference comparable to the binomial uncertainty on this sample. The median 68\% width is $0.088$ for \texttt{ANNz2}, so the neural-network posteriors are about $11\%$ narrower, but this difference does not translate into lower PIT KS or better 68\% calibration. In order to visualize the difference, figure~\ref{fig:des_pdfs} shows twelve individual examples. For the majority of galaxies the two PDFs peak at the same redshift and have comparable widths; the largest discrepancies occur for galaxies whose \texttt{MAHA-z} posterior is itself double-peaked, in which case \texttt{MAHA-z} tends to concentrate on the mode closest to the true redshift.

\begin{figure*}
    \centering
    \includegraphics[width=\linewidth]{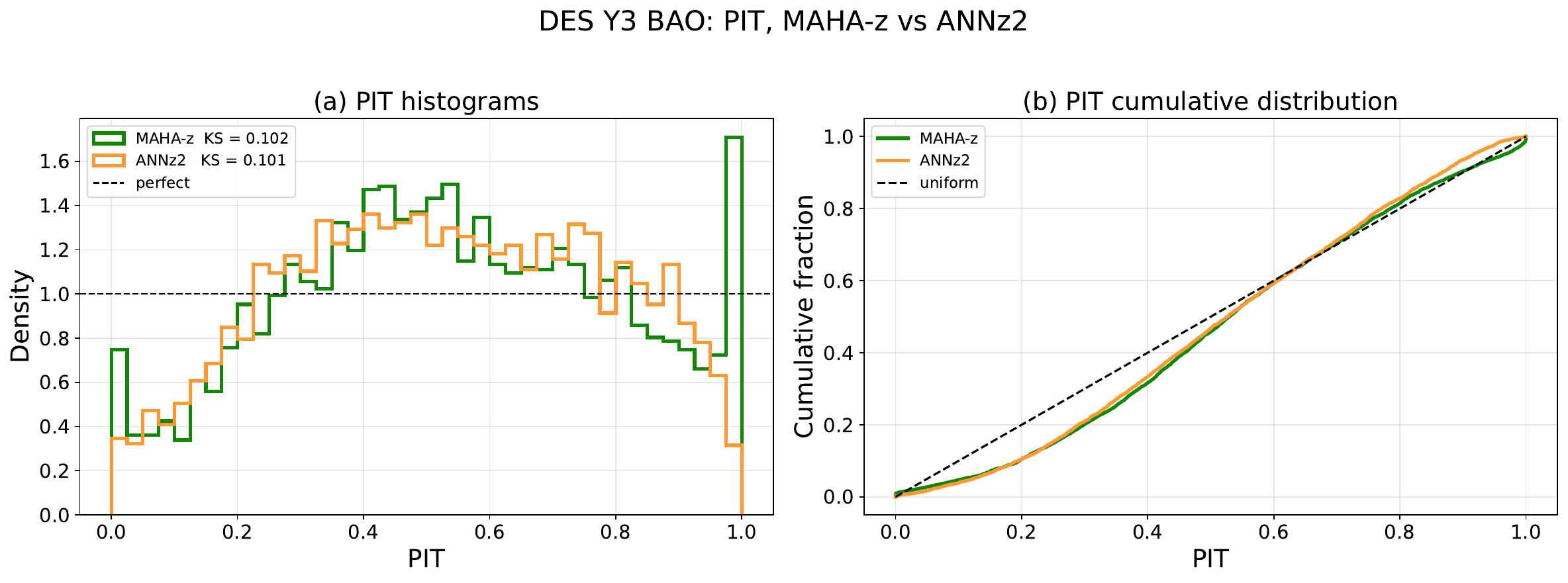}
    \caption{PIT diagnostics for \texttt{MAHA-z} (green) and \texttt{ANNz2} (orange) on the $5\,080$ galaxies common to the two DES Y3 BAO catalogues.
    \emph{(a)}~Histogram of the PIT, for each estimator. \emph{(b)}~Cumulative distribution of the same PIT values, compared to the uniform expectation. The two estimators are indistinguishable: the KS distance from uniformity is $0.101$ for \texttt{ANNz2} and $0.102$ for \texttt{MAHA-z}, both consistent with a mildly over-wide posterior that places the true redshift slightly too often near the median.}
    \label{fig:des_pit}
\end{figure*}

Figure~\ref{fig:des_pit} shows the PIT histograms for the two estimators on the same galaxies. Both distributions are close to uniform but show the characteristic mild peak near $\xi\approx0.5$ that signals a posterior wider than the true local distribution. This feature is common to the two estimators and consistent with the
68\% empirical coverages reported in Table~\ref{tab:des_vs_annz2} ($0.796$ for \texttt{ANNz2} and $0.778$ for \texttt{MAHA-z}, against a nominal $0.68$).
The KS distances from uniformity are $0.101$ for \texttt{ANNz2} and $0.102$ for \texttt{MAHA-z}, a difference of $0.001$ that is well below the
sampling noise of the $5\,080$-galaxy sample. At the 95\% level the over-coverage disappears: \texttt{MAHA-z} reaches $0.939$ and \texttt{ANNz2} reaches $0.983$, both compatible with the nominal value within the binomial uncertainty of the sample. The PIT comparison therefore confirms what the summary table already indicated: the two sets of posteriors are statistically equivalent on this dataset, and any residual difference lies in the tails of the distribution rather than in its core.

The shape classification of the \texttt{MAHA-z} posteriors is summarised as follows. Using a tolerant Gaussianity test that combines the Kolmogorov--Smirnov distance from a best-fit Gaussian ($D_{\rm KS}<0.15$), the skewness ($|s|<1$), and the excess kurtosis ($|k|<2$), and that allows one of the three criteria to fail, $63.6$~per cent of the all-colours PDFs and $66.0$~per cent of the $i$-band reference PDFs are classified as Gaussian; only $0.15$~per~cent and $0.14$~per~cent respectively are classified as skewed, and none fall into the heavy- or light-tailed classes. The remaining third of the PDFs ($36.3$~per~cent and $33.8$~per~cent) are classified as bimodal, which reflects the prevalence of genuine colour--redshift degeneracies in the photometry rather than a failure of the estimator: a $k$-neighbour KDE retains both solutions when the training set contains two distinct redshifts for the same colours, whereas a single Gaussian would average them into a broader peak centred between them. For comparison, the Pz~Cats analysis of the same DES~Y3 BAO sample reports that approximately $60$~per~cent of the \texttt{ANNz2} posteriors satisfy their Gaussian criterion \citep{Ferreira:2025mnb}. The Gaussian fraction of \texttt{MAHA-z} is therefore comparable to---and, for the $i$-band reference configuration, slightly higher than---that of a well-trained neural network on the same photometry, despite the different functional form of the two posteriors: \texttt{MAHA-z} builds its PDF from a kernel density estimate over a handful of neighbour redshifts, while \texttt{ANNz2} constructs a weighted mixture of Gaussians. The two estimators also share the same qualitative behaviour, with the bimodal fraction in each tracing the same underlying degeneracies in the DES~Y3 BAO photometry.

The implication of Table~\ref{tab:des_vs_annz2} and Fig.~\ref{fig:des_pdfs} is that \texttt{MAHA-z} produces posteriors that are as informative as those of a well-trained neural network on the same photometry, at a fraction of the computational cost. The \texttt{ANNz2} model requires around twelve hours of training on a single CPU node before any inference can be performed; \texttt{MAHA-z} requires no training, processes $7\times10^{6}$ galaxies in about ten minutes, and produces posteriors with the same calibration and comparable width. For surveys of the size of DES, LSST, or \textit{Euclid}, where the photo-$z$ catalogue must be produced, calibrated, and iterated many times during the survey, a method that gives equivalent posteriors at two orders of magnitude lower computational cost is a practical advantage.

\section{Conclusions}
\label{sec:conclusions}
In this work, we presented the photo-z estimator \texttt{MAHA-z}, it uses the Mahalanobis distance as metric for the $k$NN method. \texttt{MAHA-z} assigns each target galaxy a redshift by combining (i) a feature vector from magnitudes, colours, and optional morphology; (ii) a Mahalanobis metric with a covariance that captures the galaxy-population structure; and (iii) a $k$-nearest neighbour regression whose mean is the point estimate and whose spread is the per-galaxy uncertainty. Optionally, a KDE over the same neighbourhood yields the full posterior $p(z \mid \mathbf{f})$.

This combination of properties is not merely aesthetic: because the covariance matrix $\mathbf{C}$ absorbs the units and the correlations of every feature, no hand-tuned normalisation is required before the metric is applied, and features that would be incommensurable under a Euclidean metric---a magnitude, a colour, a S\'ersic index, an effective radius---enter Eq.~\eqref{eq:mahalanobis} on the same footing. The same property is what enables the subspace formulation of Sect.~\ref{sec:subspace}: the restriction of $\mathbf{C}$ to any subset of available features is itself a valid covariance matrix, so \texttt{MAHA-z} handles heterogeneous photometry without imputation, a situation that neural-network estimators, with their fixed input dimensionality, cannot address without an explicit in painting step.

We validated \texttt{MAHA-z} on four datasets that span the range from a controlled simulation to real survey photometry: the Euclid Flagship~2 mock, the Legacy-KiDZ calibration sample, the VIPERS survey, and the DES Y3 BAO sample. On the Euclid Flagship~2 simulation the reference LSST+NISP configuration reaches $\sigma_{\rm NMAD}=0.0148$, a catastrophic outlier fraction of $0.27\%$, and an empirical 68\% coverage of $0.933$; morphology adds a marginal improvement in the outlier tail ($f_{\rm out}=0.075\%$) for a slight penalty in scatter ($\sigma_{\rm NMAD}=0.0209$), and it halves the scatter of the NISP-only configuration ($0.101\to0.046$) precisely because near-infrared photometry alone is a poor redshift indicator and the extra morphological dimensions break part of the colour--redshift degeneracy. The same data allow us to see the metric at work at the level of individual galaxies: the fraction of multi-modal posteriors drops from $85\%$ for NISP alone to $13\%$ once optical photometry pins the 4000\,\AA\ break, which is the behaviour expected if the posterior is genuinely probing a colour--redshift degeneracy rather than merely absorbing noise.

On real data the estimator is competitive with, but not always superior to, the established codes. On KiDS it reaches $\sigma_{\rm NMAD}=0.069$ against $0.061$ for the official \texttt{BPZ} catalogue, a $13\%$ scatter penalty that is the expected price of a purely empirical estimator, but it reduces the systematic bias from $+0.018$ to
$-0.0027$, a factor of about seven, and reaches $0.061$ against $0.054$ on the bright subsample where the apertures used as morphological features are better measured. On VIPERS it attains $\sigma_{\rm NMAD}=0.039$--$0.040$ with a bias of $-0.0024$ and an outlier fraction of $2.8$--$3.7\%$, consistent with the published template-fitting estimates for the same survey and, in the case of the bias, more accurate than the $\lvert\Delta z\rvert<0.02$ tolerance adopted in earlier work. On DES Y3 it reaches
$\sigma_{\rm NMAD}\approx0.023$ with an outlier fraction below $1\%$, the best point-estimate performance among the real-data samples we considered.

The posteriors are equally informative. The PDF diagnostics we report for the Euclid, KiDS, VIPERS, and DES samples show that the KDE is mildly conservative at the $1\sigma$ level, with empirical 68\% coverage between $0.56$ and $0.93$ against a nominal $0.68$, and well calibrated at the $95\%$ level for the best-sampled cases. The
conservatism is a consequence of the small-$k$ KDE: a five-neighbour kernel cannot resolve structure finer than the local inter-neighbour spacing, so the posterior is necessarily broader than the true local redshift distribution. It can be reduced by narrowing $h_a$ in Eq.~\eqref{eq:bandwidth}, at the cost of a larger catastrophic outlier fraction, and we retain the default bandwidth because the outlier tail, not the core of the posterior, is what dominates the cosmological error budget. The most striking PDF result is the DES comparison: \texttt{MAHA-z} and \texttt{ANNz2} produce statistically indistinguishable posteriors on the same $5\,080$ galaxies, with PIT KS distances of $0.102$ and $0.101$ and 68\% coverages of $0.778$ and $0.796$, even though \texttt{ANNz2} requires around twelve hours of training on a single CPU node while \texttt{MAHA-z} processes $7\times10^{6}$ galaxies in about ten minutes with no training phase beyond the construction of the covariance matrix.

The internal structure of the \texttt{MAHA-z} posteriors confirms that the estimator is extracting physical redshift information. A redshift-resolved leave-one-band-out analysis  Fig.~\ref{fig:loo_per_zbin_heatmap} shows that each filter matters precisely where the 4000\,\AA\ break crosses it---most strikingly the $i$ band at $z\approx0.75$, where removing it alone raises the NMAD by a factor of $9.5$---while a second peak in the $u$ band at $z\approx1.8$ traces the Lyman-$\alpha$ break. The posteriors are predominantly unimodal and near-Gaussian ($63.6$ and $66.0$~per~cent of
the all-colours and $i$-reference PDFs respectively), with the remaining third almost entirely bimodal and traceable to genuine colour--redshift degeneracies; this Gaussian fraction is comparable to the $\approx60$~per~cent reported for \texttt{ANNz2} on the same sample in Pz~Cats.

In summary, \texttt{MAHA-z} offers a transparent, training-light, computationally inexpensive alternative to template-fitting and neural-network photo-$z$ codes. It does not replace them: on datasets where the training set is small or unrepresentative, and on datasets where the colour--redshift relation is highly non-linear it is matched, but not beaten, by well-trained neural networks. What it does offer is a metric that respects the geometry of the galaxy population, a posterior that is as informative as that of a neural network at a fraction of the computational cost, and a native treatment of heterogeneous photometry that neither family handles as cleanly. For the $10^{9}$--$10^{10}$ object catalogues of Stage-IV surveys, where the photo-$z$ catalogue must be produced, validated, and re-derived many times during the lifetime of the survey, these properties make \texttt{MAHA-z} a practical complement to the established codes.

\begin{acknowledgements}
PSF and RC acknowledge support from the National Key Research and Development Program of China and the Zhejiang Provincial Top-Level Research Support Program. 
RRRR thanks CNPq for partial financial support (grant no. $303829/2025-7$).

 This project used public archival data from the Dark Energy Survey (DES). Funding for the DES Projects has been provided by the U.S. Department of Energy, the U.S. National Science Foundation, the Ministry of Science and Education of Spain, the Science and Technology Facilities Council of the United Kingdom, the Higher Education Funding Council for England, the National Center for Supercomputing Applications at the University of Illinois at Urbana–Champaign, the Kavli Institute of Cosmological Physics at the University of Chicago, the Center for Cosmology and Astro-Particle Physics at the Ohio State University, the Mitchell Institute for Fundamental Physics and Astronomy at Texas A\&M University, Financiadora de Estudos e Projetos, Fundação Carlos Chagas Filho de Amparo à Pesquisa do Estado do Rio de Janeiro, Conselho Nacional de Desenvolvimento Científico e Tecnológico and the Ministério da Ciência, Tecnologia e Inovação, the Deutsche Forschungsgemeinschaft and the Collaborating Institutions in the Dark Energy Survey.

The Collaborating Institutions are Argonne National Laboratory, the University of California at Santa Cruz, the University of Cambridge, Centro de Investigaciones Enérgeticas, Medioambientales y Tecnológicas–Madrid, the University of Chicago, University College London, the DES-Brazil Consortium, the University of Edinburgh, the Eidgenössische Technische Hochschule (ETH) Zürich, Fermi National Accelerator Laboratory, the University of Illinois at Urbana-Champaign, the Institut de Ciències de l'Espai (IEEC/CSIC), the Institut de Física d'Altes Energies, Lawrence Berkeley National Laboratory, the Ludwig-Maximilians Universität München and the associated Excellence Cluster Universe, the University of Michigan, the National Optical Astronomy Observatory, the University of Nottingham, The Ohio State University, the OzDES Membership Consortium, the University of Pennsylvania, the University of Portsmouth, SLAC National Accelerator Laboratory, Stanford University, the University of Sussex, and Texas A\&M University.

Based in part on observations at Cerro Tololo Inter-American Observatory, National Optical Astronomy Observatory, which is operated by the Association of Universities for Research in Astronomy (AURA) under a cooperative agreement with the National Science Foundation. 

 This paper uses data from the VIMOS Public Extragalactic Redshift Survey (VIPERS). VIPERS has been performed using the ESO Very Large Telescope, under the ``Large Programme'' 182.A-0886. The participating institutions and funding agencies are listed at \url{http://vipers.inaf.it}.

  Based on observations made with ESO Telescopes at the La Silla Paranal Observatory under programme IDs 177.A-3016, 177.A-3017, 177.A-3018 and 179.A-2004, and on data products produced by the KiDS consortium. The KiDS production team acknowledges support from: Deutsche Forschungsgemeinschaft, ERC, NOVA and NWO-M grants; Target; the University of Padova, and the University Federico II (Naples). 

  This work has made use of CosmoHub, developed by PIC (maintained by IFAE and CIEMAT) in collaboration with ICE-CSIC. It received funding from the Spanish government (grant EQC2021-007479-P funded by MCIN/AEI/10.13039/501100011033), the EU NextGeneration/PRTR (PRTR-C17.I1), and the Generalitat de Catalunya.

\end{acknowledgements}

\bibliographystyle{aa}
\bibliography{biblio}

@article{mahalanobis,
  title={Reprint of: On the Generalised Distance in Statistics.},
  author={Mahalanobis, P.C.},
  journal={Sankhya A},
  year={2018},
  volume={80},
  pages={1 - 7},
  url={https://api.semanticscholar.org/CorpusID:239595337},
  doi={https://doi.org/10.1007/s13171-019-00164-5}
}

@article{Kids,
   title={The fifth data release of the Kilo Degree Survey: Multi-epoch optical/NIR imaging covering wide and legacy-calibration fields},
   volume={686},
   ISSN={1432-0746},
   url={http://dx.doi.org/10.1051/0004-6361/202346730},
   DOI={10.1051/0004-6361/202346730},
   journal={Astronomy \& Astrophysics},
   publisher={EDP Sciences},
   author={Wright, Angus H. and Kuijken, Konrad and Hildebrandt, Hendrik and Radovich, Mario and Bilicki, Maciej and Dvornik, Andrej and Getman, Fedor and Heymans, Catherine and Hoekstra, Henk and Li, Shun-Sheng and Miller, Lance and Napolitano, Nicola R. and Xia, Qianli and Asgari, Marika and Brescia, Massimo and Buddelmeijer, Hugo and Burger, Pierre and Castignani, Gianluca and Cavuoti, Stefano and de Jong, Jelte and Edge, Alastair and Giblin, Benjamin and Giocoli, Carlo and Harnois-Déraps, Joachim and Jalan, Priyanka and Joachimi, Benjamin and John William, Anjitha and Joudaki, Shahab and Kannawadi, Arun and Kaur, Gursharanjit and La Barbera, Francesco and Linke, Laila and Mahony, Constance and Maturi, Matteo and Moscardini, Lauro and Nakoneczny, Szymon J. and Paolillo, Maurizio and Porth, Lucas and Puddu, Emanuella and Reischke, Robert and Schneider, Peter and Sereno, Mauro and Shan, HuanYuan and Sifón, Cristóbal and Stölzner, Benjamin and Tröster, Tilman and Valentijn, Edwin and van den Busch, Jan Luca and Verdoes Kleijn, Gijs and Wittje, Anna and Yan, Ziang and Yao, Ji and Yoon, Mijin and Zhang, Yun-Hao},
   year={2024},
   month={June}, pages={A170} }

@article{Euclid:2024few,
    author = "Castander, F. J. and others",
    collaboration = "Euclid",
    title = "{Euclid - V. The Flagship galaxy mock catalogue: A comprehensive simulation for the Euclid mission}",
    eprint = "2405.13495",
    archivePrefix = "arXiv",
    primaryClass = "astro-ph.CO",
    doi = "10.1051/0004-6361/202450853",
    journal = "Astron. Astrophys.",
    volume = "697",
    pages = "A5",
    year = "2025"
}

@article{Moutard_2016I,
   title={The VIPERS Multi-Lambda Survey: I. UV and near-IR observations, multi-colour catalogues, and photometric redshifts},
   volume={590},
   ISSN={1432-0746},
   url={http://dx.doi.org/10.1051/0004-6361/201527945},
   DOI={10.1051/0004-6361/201527945},
   journal={Astronomy \& Astrophysics},
   publisher={EDP Sciences},
   author={Moutard, T. and Arnouts, S. and Ilbert, O. and Coupon, J. and Hudelot, P. and Vibert, D. and Comte, V. and Conseil, S. and Davidzon, I. and Guzzo, L. and Llebaria, A. and Martin, C. and McCracken, H. J. and Milliard, B. and Morrison, G. and Schiminovich, D. and Treyer, M. and Van Werbaeke, L.},
   year={2016},
   month=May, pages={A102} }

@ARTICLE{Kriek2011,
       author = {{Kriek}, Mariska and {van Dokkum}, Pieter G. and {Whitaker}, Katherine E. and {Labb{\'e}}, Ivo and {Franx}, Marijn and {Brammer}, Gabriel B.},
        title = "{H{\ensuremath{\alpha}} and 4000 {\r{A}} Break Measurements for \raisebox{-0.5ex}\textasciitilde3500 K-selected Galaxies at 0.5 < z < 2.0}",
      journal = {\apj},
         year = {2011},
        month = {dec},
       volume = {743},
       number = {2},
          eid = {168},
        pages = {168},
          doi = {10.1088/0004-637X/743/2/168},
archivePrefix = {arXiv},
       eprint = {1103.0279},
 primaryClass = {astro-ph.CO},
       adsurl = {https://ui.adsabs.harvard.edu/abs/2011ApJ...743..168K}
}

@article{Moutard_2016II,
   title={The VIPERS Multi-Lambda Survey: II. Diving with massive galaxies in 22 square degrees since z= 1.5},
   volume={590},
   ISSN={1432-0746},
   url={http://dx.doi.org/10.1051/0004-6361/201527294},
   DOI={10.1051/0004-6361/201527294},
   journal={Astronomy \& Astrophysics},
   publisher={EDP Sciences},
   author={Moutard, T. and Arnouts, S. and Ilbert, O. and Coupon, J. and Davidzon, I. and Guzzo, L. and Hudelot, P. and McCracken, H. J. and Van Werbaeke, L. and Morrison, G. E. and Le Fèvre, O. and Comte, V. and Bolzonella, M. and Fritz, A. and Garilli, B. and Scodeggio, M.},
   year={2016},
   month=May, pages={A103} }

@article{Raichoor_2014,
   title={THE NEXT GENERATION VIRGO CLUSTER SURVEY. XV. THE PHOTOMETRIC REDSHIFT ESTIMATION FOR BACKGROUND SOURCES},
   volume={797},
   ISSN={1538-4357},
   url={http://dx.doi.org/10.1088/0004-637X/797/2/102},
   DOI={10.1088/0004-637x/797/2/102},
   number={2},
   journal={The Astrophysical Journal},
   publisher={American Astronomical Society},
   author={Raichoor, A. and Mei, S. and Erben, T. and Hildebrandt, H. and Huertas-Company, M. and Ilbert, O. and Licitra, R. and Ball, N. M. and Boissier, S. and Boselli, A. and Chen, Y.-T. and Côté, P. and Cuillandre, J.-C. and Duc, P. A. and Durrell, P. R. and Ferrarese, L. and Guhathakurta, P. and Gwyn, S. D. J. and Kavelaars, J. J. and Lançon, A. and Liu, C. and MacArthur, L. A. and Muller, M. and Muñoz, R. P. and Peng, E. W. and Puzia, T. H. and Sawicki, M. and Toloba, E. and Van Waerbeke, L. and Woods, D. and Zhang, H.},
   year={2014},
   month=Dec, pages={102} }

@article{benitez2000bayesian,
  title={Bayesian photometric redshift estimation},
  author={Benitez, Narciso},
  journal={The Astrophysical Journal},
  volume={536},
  number={2},
  pages={571},
  year={2000},
  publisher={IOP Publishing}
}

@article{sadeh2016annz2,
  title={ANNz2: photometric redshift and probability distribution function estimation using machine learning},
  author={Sadeh, Iftach and Abdalla, Filipe B and Lahav, Ofer},
  journal={Publications of the Astronomical Society of the Pacific},
  volume={128},
  number={968},
  pages={104502},
  year={2016},
  publisher={IOP Publishing}
}

@article{Ferreira:2025mnb,
    author = "Ferreira, Paula S. and Reis, Ribamar R. R.",
    title = "{Pz Cats: photometric redshift catalogs based on DES Y3 BAO sample}",
    eprint = "2501.04118",
    archivePrefix = "arXiv",
    primaryClass = "astro-ph.GA",
    doi = "10.1088/1475-7516/2025/07/008",
    journal = "JCAP",
    volume = "07",
    pages = "008",
    year = "2025"
}

@article{scodeggio2018vimos,
  title={The VIMOS Public Extragalactic Redshift Survey (VIPERS)-Full spectroscopic data and auxiliary information release (PDR-2)},
  author={Scodeggio, MARCO and Guzzo, L and Garilli, BIANCA and Granett, BR and Bolzonella, M and De La Torre, S and Abbas, U and Adami, C and Arnouts, S and Bottini, D and others},
  journal={Astronomy \& Astrophysics},
  volume={609},
  pages={A84},
  year={2018},
  publisher={EDP Sciences}
}

@article{le2005vimos,
  title={The VIMOS VLT deep survey-First epoch VVDS-deep survey: 11 564 spectra with 17.5$\leq$ I $\leq$ 24, and the redshift distribution over 0$\leq$ z$\leq$ 5},
  author={Le Fèvre, Olivier and Vettolani, G and Garilli, B and Tresse, L and Bottini, DBVL and Le Brun, V and Maccagni, D and Picat, JP and Scaramella, R and Scodeggio, M and others},
  journal={Astronomy \& Astrophysics},
  volume={439},
  number={3},
  pages={845--862},
  year={2005},
  publisher={EDP Sciences}
}

@article{Rosell_2021,
   title={Dark Energy Survey Year 3 results: galaxy sample for BAO measurement},
   volume={509},
   ISSN={1365-2966},
   url={http://dx.doi.org/10.1093/mnras/stab2995},
   DOI={10.1093/mnras/stab2995},
   number={1},
   journal={Monthly Notices of the Royal Astronomical Society},
   publisher={Oxford University Press (OUP)},
   author={Rosell, A Carnero and Rodriguez-Monroy, M and Crocce, M and Elvin-Poole, J and Porredon, A and Ferrero, I and Mena-Fernández, J and Cawthon, R and De Vicente, J and Gaztanaga, E and Ross, A J and Sanchez, E and Sevilla-Noarbe, I and Alves, O and Andrade-Oliveira, F and Asorey, J and Avila, S and Brandao-Souza, A and Camacho, H and Chan, K C and Ferté, A and Muir, J and Riquelme, W and Rosenfeld, R and Cid, D Sanchez and Hartley, W G and Weaverdyck, N and Abbott, T and Aguena, M and Allam, S and Annis, J and Bertin, E and Brooks, D and Buckley-Geer, E and Burke, D and Calcino, J and Carollo, D and Kind, M Carrasco and Carretero, J and Castander, F and Choi, A and Conselice, C and Costanzi, M and da Costa, L and da Silva Pereira, M E and Davis, T and Desai, S and Diehl, H T and Doel, P and Drlica-Wagner, A and Eckert, K and Everett, S and Evrard, A and Flaugher, B and Fosalba, P and Frieman, J and Garcia-Bellido, J and Gerdes, D and Giannantonio, T and Glazebrook, K and Gruen, D and Gruendl, R and Gschwend, J and Gutierrez, G and Hinton, S and Hollowood, D and Honscheid, K and Hoyle, B and Huterer, D and James, D and Kim, A and Krause, E and Kuehn, K and Lahav, O and Lewis, G and Lidman, C and Lima, M and Maia, M and Malik, U and Marshall, J and Menanteau, F and Miquel, R and Mohr, J and Moller, A and Morgan, R and Ogando, R and Palmese, A and Paz-Chinchon, F and Percival, W and Pieres, A and Malagón, A Plazas and Roodman, A and Scarpine, V and Schubnell, M and Serrano, S and Sharp, R and Sheldon, E and Smith, M and Soares-Santos, M and Suchyta, E and Swanson, M and Tarle, G and Thomas, D and To, C and Tucker, B and Tucker, D and Uddin, S and Varga, T N and },
   year={2021},
   month=Oct, pages={778–799} }

@inproceedings{bartusek2022nancy,
  title={Nancy Grace Roman space telescope observatory implementation and challenges},
  author={Bartusek, Lisa ML and Davis, Jody L and Vess, Melissa F},
  booktitle={2022 IEEE Aerospace Conference (AERO)},
  pages={01--14},
  year={2022},
  organization={IEEE}
}

@techreport{abell2009lsst,
  TITLE = {{LSST Science Book, Version 2.0}},
  AUTHOR = {Abell, Paul A. and Allison, Julius and Anderson, Scott F. and Andrew, John R. and Angel, J. Roger P. and Armus, Lee and Arnett, David and Asztalos, S. J. and Axelrod, Tim S. and Bailey, Stephen and Ballantyne, D. R. and Bankert, Justin R. and Barkhouse, Wayne A. and Barr, Jeffrey D. and Barrientos, L. Felipe and Barth, Aaron J. and Bartlett, J.G. and Becker, Andrew C. and Becla, Jacek and Beers, Timothy C. and Bernstein, Joseph P. and Biswas, Rahul and Blanton, Michael R. and Bloom, Joshua S. and Bochanski, John J. and Boeshaar, Pat and Borne, Kirk D. and Bradac, Marusa and Brandt, W. N. and Bridge, Carrie R. and Brown, Michael E. and Brunner, Robert J. and Bullock, James S. and Burgasser, Adam J. and Burge, James H. and Burke, David L. and Cargile, Phillip A. and Chandrasekharan, Srinivasan and Chartas, George and Chesley, Steven R. and Chu, You-Hua and Cinabro, David and Claire, Mark W. and Claver, Charles F. and Clowe, Douglas and Connolly, A. J. and Cook, Kem H. and Cooke, Jeff and Cooray, Asantha and Covey, Kevin R. and Culliton, Christopher S. and de Jong, Roelof and de Vries, Willem H. and Debattista, Victor P. and Delgado, Francisco and Dell'Antonio, Ian P. and Dhital, Saurav and Di Stefano, Rosanne and Dickinson, Mark and Dilday, Benjamin and Djorgovski, S. G. and Dobler, Gregory and Donalek, Ciro and Dubois-Felsmann, Gregory and Durech, Josef and Eliasdottir, Ardis and Eracleous, Michael and Eyer, Laurent and Falco, Emilio E. and Fan, Xiaohui and Fassnacht, Christopher D. and Ferguson, Harry C. and Fernandez, Yanga R. and Fields, Brian D. and Finkbeiner, Douglas and Figueroa, Eduardo E. and Fox, Derek B. and Francke, Harold and Frank, James S. and Frieman, Josh and Fromenteau, S. and Furqan, Muhammad and Galaz, Gaspar and Gal-Yam, A. and Garnavich, Peter and Gawiser, Eric and Geary, John and Gee, Perry and Gibson, Robert R. and Gilmore, Kirk and Grace, Emily A. and Green, Richard F. and Gressler, William J. and Grillmair, Carl J. and Habib, Salman and Haggerty, J. S. and Hamuy, Mario and Harris, Alan W. and Hawley, Suzanne L. and Heavens, Alan F. and Hebb, Leslie and Henry, Todd J. and Hileman, Edward and Hilton, Eric J. and Hoadley, Keri and Holberg, J. B. and Holman, Matt J. and Howell, Steve B. and Infante, Leopoldo and Ivezic, Zeljko and Jacoby, Suzanne H. and Jain, Bhuvnesh and Jedicke, R. and Jee, M. James and Garrett Jernigan, J. and Jha, Saurabh W. and Johnston, Kathryn V. and Jones, R. Lynne and Juric, Mario and Kaasalainen, Mikko and Kafka, Styliani and Kahn, Steven M. and Kaib, Nathan A. and Kalirai, Jason and Kantor, Jeff and Kasliwal, Mansi M. and Keeton, Charles R. and Kessler, Richard and Knezevic, Zoran and Kowalski, Adam and Krabbendam, Victor L. and Krughoff, K. Simon and Kulkarni, Shrinivas and Kuhlman, Stephen and Lacy, Mark and Lepine, Sebastien and Liang, Ming and Lien, Amy and Lira, Paulina and Long, Knox S. and Lorenz, Suzanne and Lotz, Jennifer M. and Lupton, R. H. and Lutz, Julie and Macri, Lucas M. and Mahabal, Ashish A. and Mandelbaum, Rachel and Marshall, Phil and May, Morgan and Mcgehee, Peregrine M. and Meadows, Brian T. and Meert, Alan and Milani, Andrea and Miller, Christopher J. and Miller, Michelle and Mills, David and Minniti, Dante and Monet, David and Mukadam, Anjum S. and Nakar, Ehud and Neill, Douglas R. and Newman, Jeffrey A. and Nikolaev, Sergei and Nordby, Martin and O'Connor, Paul and Oguri, Masamune and Oliver, John and Olivier, Scot S. and Olsen, Julia K. and Olsen, Knut and Olszewski, Edward W. and Oluseyi, Hakeem and Padilla, Nelson D. and Parker, Alex and Pepper, Joshua and Peterson, John R. and Petry, Catherine and Pinto, Philip A. and Pizagno, James L. and Popescu, Bogdan and Prsa, Andrej and Radcka, Veljko and Raddick, M. Jordan and Rasmussen, Andrew and Rau, Arne and Rho, Jeonghee and Rhoads, James E. and Richards, Gordon T. and Ridgway, Stephen T. and Robertson, Brant E. and Roskar, Rok and Saha, Abhijit and Sarajedini, Ata and Scannapieco, Evan and Schalk, Terry and Schindler, Rafe and Schmidt, Samuel and Schmidt, Sarah and Schneider, Donald P. and Schumacher, German and Scranton, Ryan and Sebag, Jacques and Seppala, Lynn G. and Shemmer, Ohad and Simon, Joshua D. and Sivertz, M. and Smith, Howard A. and Allyn Smith, J. and Smith, Nathan and Spitz, Anna H. and Stanford, Adam and Stassun, Keivan G. and Strader, Jay and Strauss, Michael A. and Stubbs, Christopher W. and Sweeney, Donald W. and Szalay, Alex and Szkody, Paula and Takada, Masahiro and Thorman, Paul and Trilling, David E. and Trimble, Virginia and Tyson, Anthony and van Berg, Richard and Vanden Berk, Daniel and Vanderplas, Jake and Verde, Licia and Vrsnak, Bojan and Walkowicz, Lucianne M. and Wandelt, Benjamin D. and Wang, Sheng and Wang, Yun and Warner, Michael and Wechsler, Risa H. and West, Andrew A. and Wiecha, Oliver and Williams, Benjamin F. and Willman, Beth and Wittman, David and Wolff, Sidney C. and Wood-Vasey, W. Michael and Wozniak, Przemek and Young, Patrick and Zentner, Andrew and Zhan, Hu},
  URL = {https://in2p3.hal.science/in2p3-00711851},
  NOTE = {596 pages. Also available at full resolution at http://www.lsst.org/lsst/scibook},
  PAGES = {201},
  INSTITUTION = {{-}},
  YEAR = {2009},
  MONTH = {Dec},
  HAL_ID = {in2p3-00711851},
  HAL_VERSION = {v1},
}

@article{csst,
  title={Forecasting the BAO measurements of the CSST galaxy and AGN spectroscopic surveys},
  author={Miao, Haitao and Gong, Yan and Chen, Xuelei and Huang, Zhiqi and Li, Xiao-Dong and Zhan, Hu},
  journal={Monthly Notices of the Royal Astronomical Society},
  volume={531},
  number={4},
  pages={3991--4005},
  year={2024},
doi={https://doi.org/10.1093/mnras/stae1370},
  publisher={Oxford University Press}
}

@ARTICLE{euclid,
       author = {{Euclid Collaboration} and {Mellier}, Y. and {Abdurro'uf} and {Acevedo Barroso}, J.~A. and {Ach{\'u}carro}, A. and {Adamek}, J. and {Adam}, R. and {Addison}, G.~E. and {Aghanim}, N. and {Aguena}, M. and {Ajani}, V. and {Akrami}, Y. and {Al-Bahlawan}, A. and {Alavi}, A. and {Albuquerque}, I.~S. and {Alestas}, G. and {Alguero}, G. and {Allaoui}, A. and {Allen}, S.~W. and {Allevato}, V. and {Alonso-Tetilla}, A.~V. and {Altieri}, B. and {Alvarez-Candal}, A. and {Alvi}, S. and {Amara}, A. and {Amendola}, L. and {Amiaux}, J. and {Andika}, I.~T. and {Andreon}, S. and {Andrews}, A. and {Angora}, G. and {Angulo}, R.~E. and {Annibali}, F. and {Anselmi}, A. and {Anselmi}, S. and {Arcari}, S. and {Archidiacono}, M. and {Aric{\`o}}, G. and {Arnaud}, M. and {Arnouts}, S. and {Asgari}, M. and {Asorey}, J. and {Atayde}, L. and {Atek}, H. and {Atrio-Barandela}, F. and {Aubert}, M. and {Aubourg}, E. and {Auphan}, T. and {Auricchio}, N. and {Aussel}, B. and {Aussel}, H. and {Avelino}, P.~P. and {Avgoustidis}, A. and {Avila}, S. and {Awan}, S. and {Azzollini}, R. and {Baccigalupi}, C. and {Bachelet}, E. and {Bacon}, D. and {Baes}, M. and {Bagley}, M.~B. and {Bahr-Kalus}, B. and {Balaguera-Antolinez}, A. and {Balbinot}, E. and {Balcells}, M. and {Baldi}, M. and {Baldry}, I. and {Balestra}, A. and {Ballardini}, M. and {Ballester}, O. and {Balogh}, M. and {Ba{\~n}ados}, E. and {Barbier}, R. and {Bardelli}, S. and {Baron}, M. and {Barreiro}, T. and {Barrena}, R. and {Barriere}, J.-C. and {Barros}, B.~J. and {Barthelemy}, A. and {Bartolo}, N. and {Basset}, A. and {Battaglia}, P. and {Battisti}, A.~J. and {Baugh}, C.~M. and {Baumont}, L. and {Bazzanini}, L. and {Beaulieu}, J.-P. and {Beckmann}, V. and {Belikov}, A.~N. and {Bel}, J. and {Bellagamba}, F. and {Bella}, M. and {Bellini}, E. and {Benabed}, K. and {Bender}, R. and {Benevento}, G. and {Bennett}, C.~L. and {Benson}, K. and {Bergamini}, P. and {Bermejo-Climent}, J.~R. and {Bernardeau}, F. and {Bertacca}, D. and {Berthe}, M. and {Berthier}, J. and {Bethermin}, M. and {Beutler}, F. and {Bevillon}, C. and {Bhargava}, S. and {Bhatawdekar}, R. and {Bianchi}, D. and {Bisigello}, L. and {Biviano}, A. and {Blake}, R.~P. and {Blanchard}, A. and {Blazek}, J. and {Blot}, L. and {Bosco}, A. and {Bodendorf}, C. and {Boenke}, T. and {B{\"o}hringer}, H. and {Boldrini}, P. and {Bolzonella}, M. and {Bonchi}, A. and {Bonici}, M. and {Bonino}, D. and {Bonino}, L. and {Bonvin}, C. and {Bon}, W. and {Booth}, J.~T. and {Borgani}, S. and {Borlaff}, A.~S. and {Borsato}, E. and {Bose}, B. and {Botticella}, M.~T. and {Boucaud}, A. and {Bouche}, F. and {Boucher}, J.~S. and {Boutigny}, D. and {Bouvard}, T. and {Bouwens}, R. and {Bouy}, H. and {Bowler}, R.~A.~A. and {Bozza}, V. and {Bozzo}, E. and {Branchini}, E. and {Brando}, G. and {Brau-Nogue}, S. and {Brekke}, P. and {Bremer}, M.~N. and {Brescia}, M. and {Breton}, M.-A. and {Brinchmann}, J. and {Brinckmann}, T. and {Brockley-Blatt}, C. and {Brodwin}, M. and {Brouard}, L. and {Brown}, M.~L. and {Bruton}, S. and {Bucko}, J. and {Buddelmeijer}, H. and {Buenadicha}, G. and {Buitrago}, F. and {Burger}, P. and {Burigana}, C. and {Busillo}, V. and {Busonero}, D. and {Cabanac}, R. and {Cabayol-Garcia}, L. and {Cagliari}, M.~S. and {Caillat}, A. and {Caillat}, L. and {Calabrese}, M. and {Calabro}, A. and {Calderone}, G. and {Calura}, F. and {Camacho Quevedo}, B. and {Camera}, S. and {Campos}, L. and {Ca{\~n}as-Herrera}, G. and {Candini}, G.~P. and {Cantiello}, M. and {Capobianco}, V. and {Cappellaro}, E. and {Cappelluti}, N. and {Cappi}, A. and {Caputi}, K.~I. and {Cara}, C. and {Carbone}, C. and {Cardone}, V.~F. and {Carella}, E. and {Carlberg}, R.~G. and {Carle}, M. and {Carminati}, L. and {Caro}, F. and {Carrasco}, J.~M. and {Carretero}, J. and {Carrilho}, P. and {Carron Duque}, J. and {Carry}, B.},
        title = "{Euclid: I. Overview of the Euclid mission}",
      journal = {A\&A},
         year = {2025},
        month = {may},
       volume = {697},
          eid = {A1},
        pages = {A1},
          doi = {10.1051/0004-6361/202450810},
archivePrefix = {arXiv},
       eprint = {2405.13491},
 primaryClass = {astro-ph.CO},
       adsurl = {https://ui.adsabs.harvard.edu/abs/2025A&A...697A...1E}
}

@article{arnouts1999measuring,
  title={Measuring and modelling the redshift evolution of clustering: the Hubble Deep Field North},
  author={Arnouts, Stephane and Cristiani, Stefano and Moscardini, Lauro and Matarrese, Sabino and Lucchin, Francesco and Fontana, Adriano and Giallongo, Emanuele},
  journal={Monthly Notices of the Royal Astronomical Society},
  volume={310},
  number={2},
  pages={540--556},
  year={1999},
  publisher={Blackwell Science Ltd Oxford, UK}
}

@article{guzzo2014vimos,
  title={The VIMOS Public Extragalactic Redshift Survey (VIPERS)-An unprecedented view of galaxies and large-scale structure at 0.5< z< 1.2},
  author={Guzzo, L and Scodeggio, M and Garilli, B and Granett, BR and Fritz, A and Abbas, U and Adami, C and Arnouts, S and Bel, J and Bolzonella, M and others},
  journal={Astronomy \& Astrophysics},
  volume={566},
  pages={A108},
  year={2014},
  publisher={EDP Sciences}
}

@article{feldmann2006zurich,
  title={The zurich extragalactic bayesian redshift analyzer and its first application: COSMOS},
  author={Feldmann, Robert and Carollo, C Marcella and Porciani, C and Lilly, SJ and Capak, Peter and Taniguchi, Y and F{\`e}vre, O Le and Renzini, Alvio and Scoville, Nick and Ajiki, M and others},
  journal={Monthly Notices of the Royal Astronomical Society},
  volume={372},
  number={2},
  pages={565--577},
  year={2006},
  publisher={Blackwell Publishing Ltd Oxford, UK}
}

@article{brammer2008eazy,
  title={EAZY: a fast, public photometric redshift code},
  author={Brammer, Gabriel B and van Dokkum, Pieter G and Coppi, Paolo},
  journal={The Astrophysical Journal},
  volume={686},
  number={2},
  pages={1503},
  year={2008},
  publisher={IOP Publishing}
}

@article{10.1093/mnras/stt574,
    author = {Carrasco Kind, Matias and Brunner, Robert J.},
    title = {TPZ: photometric redshift PDFs and ancillary information by using prediction trees and random forests},
    journal = {Monthly Notices of the Royal Astronomical Society},
    volume = {432},
    number = {2},
    pages = {1483-1501},
    year = {2013},
    month = {05},
    issn = {0035-8711},
    doi = {10.1093/mnras/stt574},
    url = {https://doi.org/10.1093/mnras/stt574},
    eprint = {https://academic.oup.com/mnras/article-pdf/432/2/1483/18463634/stt574.pdf},
}

@article{de2016dnf,
  title={DNF--Galaxy photometric redshift by Directional Neighbourhood Fitting},
  author={De Vicente, Juan and S{\'a}nchez, Eusebio and Sevilla-Noarbe, Ignacio},
  journal={Monthly Notices of the Royal Astronomical Society},
  volume={459},
  number={3},
  pages={3078--3088},
  year={2016},
  publisher={Oxford University Press}
}

@article{almosallam2016gpz,
  title={GPz: non-stationary sparse Gaussian processes for heteroscedastic uncertainty estimation in photometric redshifts},
  author={Almosallam, Ibrahim A and Jarvis, Matt J and Roberts, Stephen J},
  journal={Monthly Notices of the Royal Astronomical Society},
  volume={462},
  number={1},
  pages={726--739},
  year={2016},
  publisher={The Royal Astronomical Society}
}

@article{Toribio_San_Cipriano_2024,
   title={Dark Energy Survey Deep Field photometric redshift performance and training incompleteness assessment},
   volume={686},
   ISSN={1432-0746},
   url={http://dx.doi.org/10.1051/0004-6361/202348956},
   DOI={10.1051/0004-6361/202348956},
   journal={Astronomy \& Astrophysics},
   publisher={EDP Sciences},
   author={Toribio San Cipriano, L. and De Vicente, J. and Sevilla-Noarbe, I. and Hartley, W. G. and Myles, J. and Amon, A. and Bernstein, G. M. and Choi, A. and Eckert, K. and Gruendl, R. A. and Harrison, I. and Sheldon, E. and Yanny, B. and Aguena, M. and Allam, S. S. and Alves, O. and Bacon, D. and Brooks, D. and Campos, A. and Carnero Rosell, A. and Carretero, J. and Castander, F. J. and Conselice, C. and da Costa, L. N. and Pereira, M. E. S. and Davis, T. M. and Desai, S. and Diehl, H. T. and Doel, P. and Ferrero, I. and Frieman, J. and García-Bellido, J. and Gaztañaga, E. and Giannini, G. and Hinton, S. R. and Hollowood, D. L. and Honscheid, K. and James, D. J. and Kuehn, K. and Lee, S. and Lidman, C. and Marshall, J. L. and Mena-Fernández, J. and Menanteau, F. and Miquel, R. and Palmese, A. and Pieres, A. and Plazas Malagón, A. A. and Roodman, A. and Sanchez, E. and Smith, M. and Soares-Santos, M. and Suchyta, E. and Swanson, M. E. C. and Tarle, G. and Vincenzi, M. and Weaverdyck, N. and Wiseman, P. and },
   year={2024},
   month=May, pages={A38} }

@article{Mobasher_2007,
   title={Photometric Redshifts of Galaxies in COSMOS},
   volume={172},
   ISSN={1538-4365},
   url={http://dx.doi.org/10.1086/516590},
   DOI={10.1086/516590},
   number={1},
   journal={The Astrophysical Journal Supplement Series},
   publisher={American Astronomical Society},
   author={Mobasher, B. and Capak, P. and Scoville, N. Z. and Dahlen, T. and Salvato, M. and Aussel, H. and Thompson, D. J. and Feldmann, R. and Tasca, L. and Lefevre, O. and Lilly, S. and Carollo, C. M. and Kartaltepe, J. S. and McCracken, H. and Mould, J. and Renzini, A. and Sanders, D. B. and Shopbell, P. L. and Taniguchi, Y. and Ajiki, M. and Shioya, Y. and Contini, T. and Giavalisco, M. and Ilbert, O. and Iovino, A. and Le Brun, V. and Mainieri, V. and Mignoli, M. and Scodeggio, M.},
   year={2007},
   month=Sept, pages={117–131} }

@article{Abdalla_2011,
   title={A comparison of six photometric redshift methods applied to 1.5 million luminous red galaxies: Photometric redshifts for 1.5 million LRGs},
   volume={417},
   ISSN={0035-8711},
   url={http://dx.doi.org/10.1111/j.1365-2966.2011.19375.x},
   DOI={10.1111/j.1365-2966.2011.19375.x},
   number={3},
   journal={Monthly Notices of the Royal Astronomical Society},
   publisher={Oxford University Press (OUP)},
   author={Abdalla, F. B. and Banerji, M. and Lahav, O. and Rashkov, V.},
   year={2011},
   month=Sept, pages={1891–1903} }

@misc{bolzonella2000photometricredshiftsbasedstandard,
      title={Photometric Redshifts based on standard SED fitting procedures}, 
      author={Micol Bolzonella and Joan-Marc Miralles and Roser Pello'},
      year={2000},
      eprint={astro-ph/0003380},
      archivePrefix={arXiv},
      primaryClass={astro-ph},
      url={https://arxiv.org/abs/astro-ph/0003380}, 
}

@misc{motta2026photometricclassificationquasarsdes,
      title={Photometric classification of quasars from DES and photo-$z$ estimation with Machine Learning}, 
      author={Pablo Motta and Filipe B. Abdalla and Elcio Abdalla and Gabriel S. Costa and Camila Cardoso},
      year={2026},
      eprint={2605.18218},
      archivePrefix={arXiv},
      primaryClass={astro-ph.IM},
      url={https://arxiv.org/abs/2605.18218}, 
}

@article{Reis_2012,
   title={THE SLOAN DIGITAL SKY SURVEY CO-ADD: A GALAXY PHOTOMETRIC REDSHIFT CATALOG},
   volume={747},
   ISSN={1538-4357},
   url={http://dx.doi.org/10.1088/0004-637X/747/1/59},
   DOI={10.1088/0004-637x/747/1/59},
   number={1},
   journal={The Astrophysical Journal},
   publisher={American Astronomical Society},
   author={Reis, Ribamar R. R. and Soares-Santos, Marcelle and Annis, James and Dodelson, Scott and Hao, Jiangang and Johnston, David and Kubo, Jeffrey and Lin, Huan and Seo, Hee-Jong and Simet, Melanie},
   year={2012},
   month=Feb, pages={59} }

@article{Oyaizu_2008,
doi = {10.1086/523666},
url = {https://doi.org/10.1086/523666},
year = {2008},
month = {feb},
publisher = {},
volume = {674},
number = {2},
pages = {768},
author = {Oyaizu, Hiroaki and Lima, Marcos and Cunha, Carlos E. and Lin, Huan and Frieman, Joshua and Sheldon, Erin S.},
title = {A Galaxy Photometric Redshift Catalog for the Sloan Digital Sky Survey Data Release 6},
journal = {The Astrophysical Journal}
}

@ARTICLE{Kouroumpatzakis,
       author = {{Kouroumpatzakis}, K. and {Zezas}, A. and {Kyritsis}, E. and {Salim}, S. and {Svoboda}, J.},
        title = "{Star formation rate and stellar mass calibrations based on infrared photometry and their dependence on stellar population age and extinction}",
      journal = {\aap},
         year = 2023,
        month = may,
       volume = {673},
          eid = {A16},
        pages = {A16},
          doi = {10.1051/0004-6361/202346054},
archivePrefix = {arXiv},
       eprint = {2303.10013},
 primaryClass = {astro-ph.GA},
       adsurl = {https://ui.adsabs.harvard.edu/abs/2023A&A...673A..16K}
}

\end{document}